\documentclass[a4paper,11pt]{article}

\usepackage{amsmath,amssymb}     %%一些数学符号
\usepackage{color}
\usepackage{graphicx}
\usepackage{subfigure}
\usepackage{cite}                %%是引用格式变漂亮
\usepackage{hyperref}            %%超链接必须
\usepackage{multirow,makecell}   %%表格

\numberwithin{equation}{section}   %%公式按节编号
\def \be {\begin{equation}}
\def \ee {\end{equation}}
\def \ba {\begin{array}}
\def \ea {\end{array}}
\def \bea{\begin{eqnarray}}
\def \eea{\end{eqnarray}}
\def \nn {\nonumber}

\def \a {\alpha}
\def \b {\beta}
\def \g {\gamma}

\def \G {\Gamma}
\def \d {\delta}

\def \e {\epsilon}

\def \s {\sigma}

\def \r {\rho}

\def \mA {\mathcal A}
\def \mB {\mathcal B}

\def \mD {\mathcal D}
\def \mE {\mathcal E}

\def \mH {\mathcal H}
\def \mI {\mathcal I}
\def \mJ {\mathcal J}
\def \mK {\mathcal K}
\def \mL {\mathcal L}
\def \mM {\mathcal M}
\def \mN {\mathcal N}
\def \mO {\mathcal O}
\def \mP {\mathcal P}
\def \mQ {\mathcal Q}
\def \mR {\mathcal R}
\def \mS {\mathcal S}

\def \mU {\mathcal U}
\def \mV {\mathcal V}

\def \p {\partial}
\def \f {\frac}

\def \mc {\mathcal}

\def \lt {\left}
\def \rt {\right}

\def \td {\tilde}

\def \inf {\infty}

\def \Tr {{\textrm{Tr}}}
\def \tr {{\textrm{tr}}}
\def \lag {\langle}
\def \rag {\rangle}

\begin{document}

\title{\textbf{Holographic R\'enyi entropy for CFT with $W$ symmetry}}
\author{
Bin Chen$^{1,2,3}$\footnote{bchen01@pku.edu.cn},\,
Jiang Long$^{1}$\footnote{lj301@pku.edu.cn}\,
and
Jia-ju Zhang$^{1}$\footnote{jjzhang@pku.edu.cn}
}
\date{}

\maketitle

\begin{center}
{\it
$^{1}$Department of Physics and State Key Laboratory of Nuclear Physics and Technology,\\Peking University, 5 Yiheyuan Rd, Beijing 100871, P.~R.~China\\
\vspace{2mm}
$^{2}$Collaborative Innovation Center of Quantum Matter, 5 Yiheyuan Rd, Beijing 100871, P.~R.~China\\
$^{3}$Center for High Energy Physics, Peking University, 5 Yiheyuan Rd, Beijing 100871, P.~R.~China
}
\vspace{10mm}
\end{center}

\begin{abstract}

  In this paper we investigate the holographic R\'enyi entropy of two disjoint intervals on complex plane with small cross ratio $x$ for conformal field theory with $W$ symmetry in the ground state, which could be dual to a higher spin AdS$_3$ gravity. We focus on the cases of $W_3$ and $W_4$ symmetries. In order to see the nontrivial contributions from the $W$ fields, we calculate the R\'enyi entropy in the expansion of $x$ to order $x^8$ in both the gravity and the CFT sides. In the gravity side the classical contributions to the entanglement entropy is still given by the Ryu-Takayanagi area formula under the reasonable assumption, while the 1-loop quantum corrections have to take into account of the contributions not only from massless gravitons, but also from massless higher spin fields.  In the CFT side we still use the operator product expansion of twist operators in the small interval limit, but now we need to consider the quasiprimary fields constructed from $W$ fields, besides the ones from Virasoro Verma module.  In the large central charge limit, we obtain the classical, 1-loop, 2-loop, and 3-loop parts of the R\'enyi entropy. The classical and 1-loop results in the gravity and the CFT sides are in exact match.  This confirms the higher spin gravity/CFT correspondence, and also supports the holographic computation of R\'enyi entanglement entropy, including the quantum correction, in both the AdS$_3$ gravity and the higher spin AdS$_3$ gravity.

\end{abstract}

\baselineskip 18pt
\thispagestyle{empty}

\newpage

\tableofcontents

\section{Introduction}

In this paper we continue the study of the short interval expansion of R\'enyi entropy of two-dimensional (2D) conformal field theory (CFT) initiated in \cite{Headrick:2010zt,Calabrese:2010he,Chen:2013kpa}. Let us first review the basic definition of R\'enyi entropy. The entanglement R\'enyi entropy is an important notion of  quantum systems, and it is defined as follows \cite{nielsen2010quantum,petz2008quantum}. One can divide a system into two parts, say $A$ and its complement $B$. From the density matrix $\r$ of the whole system one may obtain the reduced density matrix of $A$ by tracing over the degrees of freedom of $B$, i.e.  $\r_A=\Tr_B \r$. Then the entanglement entropy of $A$ and $B$ is defined as
\be
S_A=-\Tr_A \r_A\log \r_A.
\ee
More generally one can define the R\'enyi entropy of $A$ and $B$ as
\be
S_A^{(n)}=-\f{1}{n-1} \log \Tr_A \r_A^n.
\ee
The entanglement entropy and the R\'enyi entropy are related by $S_A=\lim_{n \to 1} S_A^{(n)}$. Moreover one may choose  two subsystems $A$ and $B$ which are not necessarily complementary to each other, and define the R\'enyi mutual information of $A$ and $B$
\be
I_{A,B}^{(n)}=S_{A}^{(n)}+S_{B}^{(n)}-S_{A\cup B}^{(n)}.
\ee
The mutual information $I_{A,B}^{(1)}$ encodes the entanglement between $A$ and $B$.

The standard way of computing the R\'enyi entropy is the replica trick\cite{Callan:1994py}, but it is usually hard to operate for a general system. Some computations in 2D CFT and higher dimensional free field theory could be found in \cite{Calabrese:2004eu,Calabrese:2009ez,Calabrese:2009qy,Headrick:2010zt,
Calabrese:2010he,Klebanov:2011uf,Rajabpour:2011pt,Cardy:2013nua,Coser:2013qda,Datta:2013hba}. For a CFT with a gravity dual one may use the AdS/CFT correspondence \cite{Maldacena:1997re,Gubser:1998bc,Witten:1998qj} to do a simpler holographic computation in the bulk gravity. For the pure AdS gravity, it was conjectured by Ryu and Takayanagi that the leading contribution of entanglement  entropy is captured by the area of a minimal surface in the bulk with asymptotic boundary ending on $A$ \cite{Ryu:2006bv,Ryu:2006ef,Nishioka:2009un,Takayanagi:2012kg}. This so-called RT area law is reminiscent of the black hole entropy, and therefore since its proposal there have been many attempts to derive this law in the gravity\cite{Fursaev:2006ih,Fursaev:2007sg,Casini:2011kv,Fursaev:2012mp}.
Very recently, the RT area law of holographic entanglement entropy has been proved in \cite{Lewkowycz:2013nqa} by taking it as a generalized gravitational entropy. However it is fair to say that for dimension $D\geq 3$ CFT, it is not certain if the holographic entanglement entropy is exactly the entanglement entropy in the field theory, as there is short of direct computations in the field theory. Nevertheless for 2D CFT, the situation is much clearer.

In 2D CFT, due to its infinite dimensional conformal symmetries, the direct field computation is feasible. One could insert the twist operators to impose the nontrivial boundary conditions  in applying the replica trick \cite{Calabrese:2004eu,Calabrese:2009qy}. As a result,  the partition function on a higher-genus Riemann surface could be recast into the correlators of twist fields on a complex plane in an orbifold CFT, which arises from the $Z_n$ replica symmetry. For a 2D CFT on complex plane the R\'enyi entropy for one interval with length $\ell$ is universal and only depends on the central charge \cite{Calabrese:2004eu}
\be
S_n=\f{c}{6} \lt( 1+\f{1}{n} \rt) \log \f{\ell}{\e},
\ee
with $\e$ being the UV cutoff. For the entanglement entropy $S_1$, it has been reproduced by the holographic computation in \cite{Ryu:2006bv,Ryu:2006ef}. For the general $S_n,n>1$, they have been reproduced successfully in \cite{Faulkner:2013yia} from the Euclidean action of corresponding gravitational configurations. For multi-interval case, the correctness of RT law and its agreement with CFT result have been proved in \cite{Faulkner:2013yia} and \cite{Hartman:2013mia}, respectively.

In AdS$_3$/CFT$_2$ correspondence, the central charge of CFT is inversely proportional to the bulk Newton constant\cite{Brown:1986nw}
\be
c=\frac{3l}{2G_N^{(3)}}
\ee
where $l$ is the radius of AdS$_3$ spacetime and $G_N^{(3)}$ is the coupling constant.
The RT area law actually captures the contribution proportional to $c$. In the large central charge limit, this is the leading contributions in CFT. For the multi-interval cases, there are subleading contributions, which could be essential in discussing the issues like mutual information. From AdS/CFT correspondence, these subleading contributions should be counted as the quantum corrections to gravitational action\cite{Headrick:2010zt,Barrella:2013wja,Faulkner:2013ana}. In particular, the quantum correction to the holographic R\'enyi entropy has been studied in  \cite{Barrella:2013wja}. In the small interval limit with a small cross ratio $x$ on complex plane, the quantum 1-loop correction of graviton  to the R\'enyi mutual information has been calculated to order $x^8$.
According to the AdS/CFT correspondence, the graviton in the bulk corresponds to the stress tensor in the CFT side. To account the graviton correction, it is only necessary to consider the Virasoro Verma module in CFT. Actually from the study of quantum gravity in AdS$_3$, it has been known that the pure gravity partition function could be reproduced from the Virasoro Verma module\cite{Maloney:2007ud}. In \cite{Chen:2013kpa} by studying the OPE of twist operators at small interval limit, the R\'enyi mutual information of two interval  has been computed in the CFT side.  The quantum 1-loop correction of graviton to the R\'enyi mutual information in \cite{Barrella:2013wja} has been confirmed  to order $x^6$. This strongly support the holographic computation of  R\'enyi entropy beyond the classical level.

In this paper we continue the work initiated in \cite{Chen:2013kpa}. In this work we would like to consider the CFT with $W$ symmetry in its ground state. In this case, the dual gravity could be a higher spin (HS) AdS$_3$ gravity theory\cite{Campoleoni:2010zq,Henneaux:2010xg}. Therefore our investigations may not only shed light on the holographic R\'enyi entropy with matter coupling, but also allows us to understand the HS/CFT correspondence from a new angle. We focus on the Renyi mutual information of two disjoint intervals with small cross ratio. In order to see the contributions from $W$ fields with conformal weights 3 and 4 clearly, we need to find the contributions up to order $x^8$.
Using the method in \cite{Barrella:2013wja}, we calculate the quantum 1-loop contributions of spin-3 and spin-4 fields to the R\'enyi mutual information of the two interval case to order $x^8$. On the gravity side, under the assumption the classical configurations are still the handlebody geometries constructed in \cite{Faulkner:2013yia}, we consider the 1-loop fluctuations around these configurations.
One feature of the calculation is that the contributions of different spin fields are additive, and so we can calculate them separately. On the CFT side, we firstly consider only the contributions from the Virasoro Verma module and verify the results in \cite{Barrella:2013wja} to order $x^8$.
Then we consider  the extra contributions from $W$ fields in the CFT side. The leading contribution of the $W_m$ field to the R\'enyi mutual information in the CFT is of order $x^{2m}$. This can be got easily, and matches the bulk result. The subleading contributions are more difficult to get.
Different from the calculation in the gravity side, in the CFT side the contributions of $W$ fields cannot be considered separately, as they are involved with the stress tensor.  We verify that the contributions of $W_3$ field and/or $W_4$ field to order $x^8$ match exactly with the gravity results as well. This shows that the holographic prescription of computing the entanglement R\'enyi entropy for the ground state CFT not only applies to an ordinary CFT but also to CFTs with $W$ symmetries.

The remaining of the paper is arranged as follows. In Section~\ref{s2} we calculate the classical R\'enyi entropy as well as the 1-loop corrections of spin-3 and spin-4  fluctuations to order $x^8$ in the gravity side. In Section~\ref{s3} we confirm these results in the CFT side using the short interval expansion of the R\'enyi entropy. We end with conclusion and discussion in Section~\ref{s4}. Some  details of the computation in Section~\ref{s3} are presented in the Appendixes~\ref{sa}, \ref{sb}, and \ref{sc}. The Appendix~\ref{sd} is a byproduct of the paper and it is about the case of one short interval on cylinder. Some useful summation formulas are collected in Appendix~\ref{se}.

{\bf Note added} The same day this paper appeared in arXiv, there appeared another paper \cite{Perlmutter:2013paa} which has some overlaps with our work.

\section{Holographic R\'enyi entropy} \label{s2}

In this section we calculate the classical and 1-loop parts of the holographic R\'enyi entropy for two intervals with small cross ratio in CFT with $W$ symmetry. As we are considering the entanglement entropy in the vacuum state of CFT, we focus on the AdS$_3$ vacuum. The gravitational configurations are the same as the ones worked out in \cite{Faulkner:2013yia,Barrella:2013wja}. Therefore, we assume that the classical part coming from bulk gravitational action is invariant.\footnote{For more careful justification of this assumption  in  higher spin gravity, see \cite{Perlmutter:2013paa}.} But we have not only the massless  boundary gravitons but also massless higher spin fluctuations, all of which contribute to 1-loop quantum correction. As the massless field with higher spin $m$ corresponds to a pair of holomorphic and antiholomorphic operators with conformal weight $m$, their contributions to the R\'enyi entropy is of order $x^{2m}$. In order to read the nontrivial information, we need to do higher order expansion of $x$. In this work, we manage to work out the contributions to order $x^8$, which allows us to discuss the higher spin gravity to spin-4 field.

The higher spin gravity has been under active study in the past few years. For the pure spin-3 AdS$_3$ gravity, it was proposed in \cite{Campoleoni:2010zq} that its action could be written in terms of  Chern-Simons form with gauge group $SL(3,R)$. Moreover, as its asymptotic symmetry group has $W_3$ symmetry, it was conjectured that the spin-3 AdS$_3$ gravity could be holographically dual to a conformal field theory with $W_3$ symmetry but the same central charge. More generally, the Chern-Simons gravity could be defined with other Lie groups, describing the interaction of higher spin fields with gravity. When the gauge group is $SL(4,R)$, it describe both the spin-4 and spin-3 fields interacting with the gravity, which is dual to a CFT with $W(2,3,4)$ symmetry. But one may obtain only the spin-4 field interacting with the gravity by choosing the gauge group to be $SO(5)$ or $Sp(4)$ \cite{Chen:2012pc}. This truncated spin-4 gravity is conjectured to be dual to a CFT with $W(2,4)$ symmetry. In all these cases, the dual CFT has the same central charge as the one for pure AdS$_3$ gravity, so that all the higher spin fields could be set to vanish without spoiling the underlying correspondence. In other words, all the classical gravitational configurations in \cite{Barrella:2013wja} are still the classical solutions of higher spin gravity and moreover their bulk classical actions would not be changed by the presence of higher spin fields.

\subsection{The classical part}

The classical R\'enyi entropy for a 2D CFT with large central charge could be calculated using the method proposed in \cite{Hartman:2013mia,Faulkner:2013yia}. It  reduces to the monodromy problem of an ordinary differential equation.  In \cite{Barrella:2013wja} the method was used to calculate the small cross ratio expansion of the classical R\'enyi entropy for the two intervals case. Here we adopt the same prescription but only give the outline without much details. One could find the details of the calculation in \cite{Barrella:2013wja}.

We choose the subregion $A=(-\inf,-1]\cup[-y,y]\cup[1,+\inf)$ with $y$ being small, and have the cross ratio
\be
x=\f{4y}{(1+y)^2},
\ee
which is small too. The classical R\'enyi entropy could be obtained using
\be
\f{\p S_n^{cl}}{\p y}=-\f{c n}{3(n-1)} \g_3,
\ee
with
\bea
&& \g_3=-\frac{n^2-1}{2 n^2 y}+\frac{2 \left(n^2-1\right)^2 y}{3 n^4}+\frac{2 \left(n^2-1\right)^2 \left(49 n^4-2 n^2-11\right) y^3}{135 n^8}  \nn\\
&& \phantom{\g_3=}
+\frac{2 \left(n^2-1\right)^2 \left(3211 n^8-172 n^6-1056 n^4-172 n^2+376\right) y^5}{8505 n^{12}}  \nn\\
&& \phantom{\g_3=}+\frac{2\left(n^2-1\right)^2 y^7}{1913625 n^{16}}\left(740087 n^{12}-44106 n^{10}-290847 n^8-78748 n^6+149973 n^4 \rt. \nn\\
&&  \phantom{\g_3=} \lt. +67854 n^2-58213\right)+{O} (y^9).
\eea
And then we can get the classical R\'enyi mutual information
\bea \label{incl}
&& I_n^{cl}=\f{c(n+1)}{6n}\log y-S_n^{cl}  \nn\\
&& \phantom{I_n^{cl}}=\frac{c (n-1) (n+1)^2 y^2}{9 n^3}+\frac{c (n-1) (n+1)^2 \left(49 n^4-2 n^2-11\right) y^4}{810 n^7}  \nn\\
&& \phantom{I_n^{cl}=}+\frac{c (n-1) (n+1)^2 \left(3211 n^8-172 n^6-1056 n^4-172 n^2+376\right) y^6}{76545 n^{11}}  \nn\\
&& \phantom{I_n^{cl}=}+\frac{c (n-1) (n+1)^2 y^8}{22963500 n^{15}}\left(740087 n^{12}-44106 n^{10}-290847 n^8-78748 n^6+149973 n^4  \rt.  \nn\\
&&  \phantom{I_n^{cl}=} \lt. +67854 n^2-58213\right)+{O}(y^{10}).
\eea
Obviously, when $n=1$, the classical part of the mutual information in this case is vanishing, and this matches the RT area formula \cite{Headrick:2010zt}.

\subsection{The 1-loop correction}

The method of calculating the 1-loop correction of the R\'enyi entropy was given in \cite{Barrella:2013wja}, and it is directly related to the computation of the partition function \cite{Yin:2007gv,Giombi:2008vd}. The relation is that
\be
S_n^{1-loop}=-\f{1}{n-1} \lt( \log Z_n^{1-loop} - n \log Z_1^{1-loop} \rt).
\ee
For the two intervals on complex plane case, we have $Z_1^{1-loop}=1$ and the second term of the right hand side of the above equation can be omitted.

\subsubsection{Graviton}

The partition function depends on the matter content of the gravity as well as the spacetime configuration. When the spacetime is the quotient of global AdS$_3$ by a Schottky group $\G$, the 1-loop partition function for the graviton is \cite{Giombi:2008vd}
\be
Z^{1-loop}=\prod_{\g\in\mP}\prod_{m=2}^\inf \f{1}{|1-q_\g^m|},
\ee
with $\mP$ being a set of representatives of the primitive conjugacy classes of $\G$. Using this method, the small cross ratio $x$ expansion of the 1-loop correction of R\'enyi mutual information to order $x^8$ has been given in \cite{Barrella:2013wja}, and it has been confirmed to order $x^6$ from CFT computation  by considering only the contributions from the Virasoro Verma module \cite{Chen:2013kpa}.

\subsubsection{Spin-3 field}

In the presence of higher spin fields, the 1-loop partition function becomes \cite{Gaberdiel:2011zw}
\be
Z^{1-loop}=\prod_{\g\in\mP}\prod_{s}\prod_{m=s}^\inf \f{1}{|1-q_\g^m|}.
\ee
Here the product over $s$ is with respect to the spins of massless fields. For $s=2$ it reduces to the graviton case, and for $s\geq3$ it corresponds to the higher spin fields. One feature of this formula is that the contributions of the fields with different spins to the R\'enyi entropy could be separated. Therefore, the contribution from the gravitons is the same as the one got in \cite{Barrella:2013wja}. For the spin-3 field, the contribution to the R\'enyi mutual information is
\bea \label{ins3}
&& I_{n, spin-3}^{1-loop}=\frac{n }{n-1}\left(\frac{f_6 x^6}{4096 n^{12}}+\frac{3\left(f_7+(n^2-1)f_6\right)x^7}{4096 n^{14}}  \rt.   \\
&& \phantom{I_{n, spin-3}^{1-loop}=} \lt. +\frac{\left(273 f_8+504(n^2-1)f_7+2(137 n^4-250 n^2+113 )f_6\right)x^8}{196608 n^{16}}+O(x)^9\right). \nn
%&& \phantom{I_{n, spin-3}^{1-loop}}=\frac{(n+1) x^6}{2615348736000 n^{11}}\left(1382 n^{10}+28682 n^8+307961 n^6+2295661 n^4  \rt.  \nn\\
%&& \phantom{I_{n, spin-3}^{1-loop}=} \lt.+13803157 n^2+92427157\right)   \\
%&& \phantom{I_{n, spin-3}^{1-loop}=}+\frac{(n+1) x^7}{2615348736000 n^{13}}\left(4566 n^{12}+91994 n^{10}+962591 n^8+7021787 n^6  \rt.  \nn\\
%&& \phantom{I_{n, spin-3}^{1-loop}=} \lt.+41507095 n^4+275328319 n^2-20097152\right)\nn\\
%&& \phantom{I_{n, spin-3}^{1-loop}=}+\frac{(n+1) x^8}{32011868528640000 n^{15}}\left(115540581 n^{14}+2269217541 n^{12}+23229560627 n^{10}  \rt.  \nn\\
%&& \phantom{I_{n, spin-3}^{1-loop}=} \lt.+166406890247 n^8+969932798003 n^6+6383136915883 n^4-786856255211 n^2  \rt.  \nn\\
%&& \phantom{I_{n, spin-3}^{1-loop}=} \lt.-100983339671\right)+{O}(x^9). \nn
\eea
The functions $f_m$'s are defined in (\ref{fm}), and the explicit form of $I_{n, spin-3}^{1-loop}$ could be got easily using (\ref{fm18}). Note that the singular behavior in $I_{n, spin-3}^{1-loop}$ at $n=1$ is superficial. Actually, the $(n-1)$ factor in the denominator is cancelled by the $(n^2-1)$ factor in the $f_m$'s.

\subsubsection{Spin-4 field}

For the spin-4 field,  the contribution is
\bea \label{ins4}
&& I_{n, spin-4}^{1-loop}=\frac{n}{n-1} \left(\frac{f_8 x^8}{65536 n^{16}}+O(x)^9\right).
%&& \phantom{I_{n, spin-4}^{1-loop}}=\frac{(n+1) x^8}{32011868528640000 n^{15}}\left(10851 n^{14}+296451 n^{12}+4149467 n^{10}+39686267 n^8 \rt.  \nn\\
%&& \phantom{I_{n, spin-4}^{1-loop}=} \lt. +292184513 n^6+1777658113 n^4+9611679169 n^2+61430943169\right)+{O}(x^9). \nn
\eea

\section{R\'enyi entropy in the CFT side} \label{s3}

In this section we compute the short interval expansion of the R\'enyi entropy in the CFT side, and we mainly focus on the case of two intervals on complex plane with small cross ratio. As we are considering the CFT with $W$ symmetry, we must take into account quasi-primary fields constructed from the primary $W$ operators.

\subsection{OPE of twist operators}

The short interval expansion of R\'enyi entropy has been discussed intensively in \cite{Headrick:2010zt,Calabrese:2010he,Chen:2013kpa}, and here we only outline the key points.

A systematic way of computing the $n$-th R\'enyi entropy of $N$ disjoint intervals for a general 2D CFT on complex plane was proposed in \cite{Calabrese:2004eu}. The replica trick requires us to compute the partition function of the CFT on the Riemann surface $\mc R_{n,N}$ of genus $(n-1)(N-1)$, which originates from the sewing of $n$ copies of the complex plane with branch cuts. Alternatively, one may work with $n$ copies of the original CFT on a complex plane but with nontrivial boundary conditions relating the fields of different replicas at the branch cuts. Note that in the first picture there is one copy of the original CFT on $\mc R_{n,N}$ which is an orbifold CFT, and in the second picture there is one copy of the complex plane but the CFT consists of $n$ copies of the original CFT and so will be denoted by $CFT^n$. In the second picture the boundary conditions could be accounted by inserting the twist operators $\s(z,\bar z)$, $\td \s(z,\bar z)$  at the branch points which are the boundaries of the intervals. The twist operators are primary operators with conformal weights
\be h=\bar h=\f{c}{24} \lt( n-\f{1}{n} \rt),\ee
with $c$ being the central charge of the original CFT.

In the first picture we denote a local operator as $\phi(z_j,\bar z_j)$ with $z_j$ being the coordinate of $j$-th copy of the plane, and in the second picture we denote a local operator as $\phi_j(z,\bar z)$ with $\phi_j$ being an operator of the $j$-th copy of the original CFT and $z$ being the coordinate of the complex plane. The operators in the two pictures can be converted freely, and sometimes we just mix the two kinds of symbols for simplicity. For example, we take
\be
T(z_{j_1},\bar z_{j_1})T(z_{j_2},\bar z_{j_2}) \equiv T_{j_1}(z,\bar z)T_{j_2}(z,\bar z), ~~~ j_1 \neq j_2,
\ee
with the left side being an operator in the first picture, which is nonlocal, and the right side being an operator in the second picture, which is local, but the operators are taken to be different.

If we choose $A=[z_1,z_2]\cup\cdots\cup[z_{2N-1},z_{2N}]$, we have
\be
\Tr \r_A^n=
\lag \s(z_{2N},\bar z_{2N})\td \s(z_{2N-1},\bar z_{2N-1}) \cdots \s(z_{2},\bar z_{2})\td \s(z_{1},\bar z_{1}) \rag_{C}.
\ee
When $N=1$ and $A=[0,\ell]$, we have
\be \label{e2}
\Tr \r_A^n=\lag \s(\ell,\ell)\td \s(0,0) \rag_{C}=c_n \ell^{-\f{c}{6}\lt( n-\f{1}{n} \rt)},
\ee
with $c_n$ being a constant related to the normalization of the twist operators.
Then the R\'enyi entropy for one interval could be found \cite{Calabrese:2004eu}
\be
S_n=\f{c}{6}\lt( 1+\f{1}{n} \rt)\log\f{\ell}{\e},
\ee
with $\e$ being the UV cutoff.

We denote the linearly independent quasiprimary operators in $CFT^n$ as $\Phi_K(z,\bar z)$ with the conformal wights $h_K$ and $\bar h_K$. Moreover we orthogonalize these operators as
\be \label{e1}
\lag \Phi_K(z,\bar z) \Phi_L(w,\bar w) \rag_C=\f{\d_{KL}\a_K}{(z-w)^{2h_K}(\bar z-\bar w)^{2\bar h_K}}.
\ee
The product of quasiprimary operators in each replica forms a quasiprimary operator of $CFT^n$,
\be
\Phi_K(z,\bar z)=\prod_{j=0}^{n-1} \phi_{k_j}(z,\bar z).
\ee
In this case
\be
K=\{k_j\}, ~~~ \a_K=\prod_{j=0}^{n-1}\a_{k_j}, ~~~ h_K=\sum_{j=0}^{n-1} h_{k_j}, ~~~ \bar h_K=\sum_{j=0}^{n-1} \bar h_{k_j}.
\ee
We should bear in mind that not all of the quasiprimary operators can be written in this form \cite{Headrick:2010zt,Chen:2013kpa}.

When the intervals are short, at each interval we have the OPE of the twist operators in $CFT^n$
\be
\s(z,\bar z)\td \s(0,0)
=c_n \sum_K d_K \sum_{m,r\geq0} \f{a_K^m}{m!}\f{\bar a_K^r}{r!}\f{1}{z^{2h-h_K-m}\bar z^{2\bar h-\bar h_K-r}} \p^m \bar \p^r \Phi_K(0,0),
\ee
with the summation $K$ being over all the independent quasiprimary operators of $CFT^n$. Here
\bea
&& a_K^m\equiv \f{C_{h_K+m-1}^m}{C_{2h_K+m-1}^m}, ~~~ \bar a_K^r\equiv\f{C_{\bar h_K+r-1}^r}{C_{2\bar h_K+r-1}^r},
\eea
with the binomial coefficient being $C_x^y=\f{\G(x+1)}{\G(y+1)\G(x-y+1)}$. To calculate the $d_K$'s, we may just consider the one interval case $N=1$, and find that \cite{Calabrese:2010he,Chen:2013kpa}
\be \label{dk}
d_K=\f{1}{\a_K\ell^{h_K+\bar h_K}} \lim_{z\to\inf}z^{2 h_K}\bar z^{2\bar h_K}\lag \Phi_K(z,\bar z) \rag_{\mc R_{n,1}},
\ee
with $\a_K$ being a normalization coefficient in (\ref{e1}) and $\ell$ being the length of the interval in (\ref{e2}).

To use the OPE of the twist operators, we have to find the quasiprimary operators level by level. As the holomorphic and anti-holomorphic sector are decoupled and similar, we may just focus on the ones in the holomorphic sector. These operators are listed in Appendix~\ref{sa}. The OPE coefficients $d_K$ for these operators are calculated in Appendix~\ref{sb}.

We would like to consider the case of two short disjoint intervals on the complex plane. We choose $A=[0,\ell]\cup[1,1+\ell]$ with $\ell$ being small, and thus the cross ratio $x=\ell^2$ is small too. As shown in \cite{Chen:2013kpa}, the partition function of $CFT^n$ is
\bea
&& \Tr \r^n_A=c_n^2 \ell^{-\f{c}{3}\lt( n-\f{1}{n} \rt)} \lt( \sum_K \a_K d_K^2 \ell^{2h_K}
       \sum_{m,p\geq0}(-)^{m}\f{(m+p)!}{m!p!}a_K^m a_K^p
        C_{2h_K+m+p-1}^{m+p} \ell^{m+p} \rt)^2 \nn\\
&&\phantom{\Tr \r^n_A}
=c_n^2 x^{-\f{c}{6}\lt( n-\f{1}{n} \rt)} \lt( \sum_K \a_K d_K^2 x^{h_K} F(h_K,h_K;2h_K;x) \rt)^2,
\eea
with $K$ being the summation over all holomorphic quasiprimary operators and $F(h_K,h_K;2h_K;x)$ being the hypergeometric function. Note that we have used the formula
\be
\sum_{m,p\geq0}(-)^{m}\f{(m+p)!}{m!p!}a_K^m a_K^p C_{2h_K+m+p-1}^{m+p} \ell^{m+p}=F(h_K,h_K;2h_K;\ell^2).
\ee

\subsection{Ordinary CFT} \label{s3.2}

Firstly we only consider the quasiprimary operators constructed solely by the operators in the Virasoro Verma module. The process in straightforward but tedious,. Some intermediate steps in the calculation can be found in Appendix~\ref{sc}.

We obtain the R\'enyi mutual information
\be \label{in2}
I_n^{(2)}=I_n^{(2)tree}+I_n^{(2)1-loop}+I_n^{(2)2-loop}+I_n^{(2)3-loop}+\cdots.
\ee
Here we use the upper symbol $(2)$ to remind that this is the contribution from the Virasoro Verma module. The tree part, or the so-called classical part, being proportional to the central charge $c$, is
\bea \label{t0} \label{w0}
&& I_n^{(2)tree}=\frac{c (n-1) (n+1)^2 x^2}{144 n^3}+\frac{c (n-1) (n+1)^2 x^3}{144 n^3}+\frac{c (n-1) (n+1)^2 \left(1309 n^4-2 n^2-11\right) x^4}{207360 n^7}  \nn\\
&& \phantom{I_n^{(2)tree}=}
+\frac{c (n-1) (n+1)^2 \left(589 n^4-2 n^2-11\right) x^5}{103680 n^7}\nn\\
&& \phantom{I_n^{(2)tree}=}
+\frac{c (n-1) (n+1)^2 \left(805139 n^8-4244 n^6-23397 n^4-86 n^2+188\right) x^6}{156764160 n^{11}} \nn\\
&& \phantom{I_n^{(2)tree}=}
+\frac{c (n-1) (n+1)^2 \left(244439 n^8-1724 n^6-9537 n^4-86 n^2+188\right) x^7}{52254720 n^{11}} \nn\\
&& \phantom{I_n^{(2)tree}=}
+\frac{c (n-1) (n+1)^2 x^8}{1504935936000 n^{15}}
\left(6459666587 n^{12}-56285106 n^{10}-312586347 n^8-4722748 n^6 \rt. \nn\\
&&\phantom{I_n^{(2)tree}=}
\lt. +10301973 n^4+67854 n^2-58213\right) +{O}(x^9).
\eea
This is just (\ref{incl}) and matches the result in \cite{Headrick:2010zt,Hartman:2013mia,Faulkner:2013yia}. The quantum 1-1oop part, being proportional to $c^0$, is
\bea \label{t1}
&& I_n^{(2)1-loop}=\frac{(n+1) \left(n^2+11\right) \left(3 n^4+10 n^2+227\right) x^4}{3628800 n^7} \nn\\
&& \phantom{I_n^{(2)1-loop}=}
+\frac{(n+1) \left(109 n^8+1495 n^6+11307 n^4+81905 n^2-8416\right) x^5}{59875200 n^9}\nn\\
&& \phantom{I_n^{(2)1-loop}=}
+\frac{(n+1) x^6}{523069747200 n^{11}}\left(1444050 n^{10}+19112974 n^8+140565305 n^6+1000527837 n^4 \rt. \nn\\
&& \phantom{I_n^{(2)1-loop}=}\lt.-167731255 n^2-14142911\right) \nn\\
&& \phantom{I_n^{(2)1-loop}=}
+\frac{(n+1) x^7}{1569209241600 n^{13}}\left(5631890 n^{12}+72352658 n^{10}+520073477 n^8+3649714849 n^6 \rt. \nn\\
&& \phantom{I_n^{(2)1-loop}=}\lt.-767668979 n^4-140870807 n^2+13778112\right) \nn\\
&& \phantom{I_n^{(2)1-loop}=}
+\frac{(n+1)x^8}{3766102179840000 n^{15}} \left(16193555193 n^{14}+202784829113 n^{12}+1429840752361 n^{10} \rt. \nn\\
&& \phantom{I_n^{(2)1-loop}=}\lt.+9916221391201 n^8-2370325526301 n^6-689741905741 n^4 \rt. \nn\\
&& \phantom{I_n^{(2)1-loop}=}\lt.+59604098747 n^2+161961045427\right)
+{O}(x^9)
\eea
 which matches the result in\cite{Barrella:2013wja} to order $x^8$. There are also the quantum 2-loop contributions, being proportional to $1/c$,
\bea \label{t2}
&& I_n^{(2)2-loop}=\frac{(n+1) \lt(n^2-4\rt) \left(19 n^8+875 n^6+22317 n^4+505625 n^2+5691964\right) x^{6}}{70053984000 n^{11} c} \nn\\
&& \phantom{I_n^{(2)2-loop}=}
+\frac{(n+1) \lt(n^2-4\rt) x^7}{326918592000 n^{13}c}\left(276 n^{10}+12571 n^8+317643 n^6+7151253 n^4\rt. \nn\\
&& \phantom{I_n^{(2)2-loop}=}\lt.+79361381 n^2-9428724\right)    \\
&& \phantom{I_n^{(2)2-loop}=}
+\frac{(n+1) \lt(n^2-4\rt)x^8}{8002967132160000 n^{15} c} \left(13294413 n^{12}+599852505 n^{10}+15034412216 n^8 \rt. \nn\\
&& \phantom{I_n^{(2)2-loop}=}\lt.+336566965090 n^6+3691373772429 n^4-744486199595 n^2-466025361058\right)+{O}(x^9), \nn
\eea
and the quantum 3-loop contributions, being proportional to $1/c^2$,
\bea \label{t3}
&& I_n^{(2)3-loop}=\frac{ (n+1)(n^2-4)(n^2-9) x^8}{111152321280000 n^{15}c^2 }\left(21 n^{10}+1994 n^8+105648 n^6+4785522 n^4 \rt. \nn\\
&& \phantom{I_n^{(2)3-loop}=}\lt.+141534331 n^2+2127620484\right)+{O}(x)^9.
\eea

\subsection{CFT with $W(2,3)$ symmetry}

Taking the contributions only from the $W_3$ field into account, we get the mutual information
\be \label{in23}
I_n^{(2,3)}=I_n^{(2,3)tree}+I_n^{(2,3)1-loop}+I_n^{(2,3)2-loop}+I_n^{(2,3)3-loop}+ \cdots,
\ee
where $I_n^{(2,3)tree}$ being the same as (\ref{t0}),
\bea \label{w1}
&& I_n^{(2,3)1-loop}= \cdots +\frac{(n+1)x^6 }{1307674368000 n^{11}} \left(3610816 n^{10}+47796776 n^8+351567243 n^6 \rt. \nn\\
&& \phantom{I_n^{(2,3)1-loop}=}\lt.+2502467423 n^4-412426559 n^2+10856301\right)  \nn\\
&& \phantom{I_n^{(2,3)1-loop}=}
+\frac{(n+1) x^7}{3923023104000 n^{13}} \left(14086574 n^{12}+181019636 n^{10}+1301627579 n^8 \rt. \nn\\
&& \phantom{I_n^{(2,3)1-loop}=}\lt.+9134819803 n^6-1856911805 n^4+60815461 n^2+4299552\right)\nn\\
&& \phantom{I_n^{(2,3)1-loop}=}
+\frac{(n+1) x^8}{64023737057280000 n^{15}}\left(275521519443 n^{14}+3451880530003 n^{12} \rt. \nn\\
&& \phantom{I_n^{(2,3)1-loop}=}\lt.+24353751911391 n^{10}+168908577430911 n^8-38355668351111 n^6 \rt. \nn\\
&& \phantom{I_n^{(2,3)1-loop}=}\lt.+1040661434169 n^4-560442831723 n^2+2551371092917\right)
+{O}(x^9),
\eea
with the $\cdots$ being the $x^4$, $x^5$ parts of $I_n^{(2)1-loop}$,
\bea \label{w2}
&& I_n^{(2,3)2-loop}= \cdots+\frac{(n+1)(n^2-4) x^8}{4001483566080000 c n^{15}}\left(6650397 n^{12}+300093675 n^{10}+7521887359 n^8 \rt. \nn\\
&& \phantom{I_n^{(2,3)2-loop}=}\lt.+168383511650 n^6+1847152960191 n^4-354725220325 n^2-75741834947\right) \nn\\
&& \phantom{I_n^{(2,3)2-loop}=}
+{O}(x^9),
\eea
with the $\cdots$ being the $x^6$, $x^7$ parts of $I_n^{(2)2-loop}$,
and $I_n^{(2,3)3-loop}$ being the same as $I_n^{(2)3-loop}$.

Regarding the above results, several remarks are in order
\begin{itemize}
\item There is no further contributions from the $W$-generated operators to the tree-level result. This is consistent with the fact that there is no higher spin contribution to the classical action in the bulk.
\item The 1-loop result $I_n^{(2,3)1-loop}$ (\ref{w1}) is just the summation of $I_n^{(2)1-loop}$ (\ref{t1}) and $I_n^{spin-3,1-loop}$ (\ref{ins3}), and  is in exact match with holographic computation.
\item The 2-loop and 3-loop results are vanishing for $n=2$. This is in accord with the fact that the genus-1 partition function is 1-loop exact. Correspondingly the quantum correction to the gravitational configuration with the asymptotic boundary being genus-1 is 1-loop exact\cite{Maloney:2007ud}.
\item The 2-loop corrections are generically nonvanishing for $n\geq 3$. This fact is in accord with the fact that in gravity there exist higher loop corrections to the gravitational configurations with the asymptotic boundary being higher genus\cite{Yin:2007gv}.
\item The 3-loop correction seems to be vanishing at $n=3$. It is not clear if there is a good reason for this fact.
%\item We conjecture that the $m$-loop correction with $m\geq2$ is vanishing at $n=2,3,\cdots,m$.
\end{itemize}

\subsection{CFT with $W(2,4)$ symmetry}

Taking the contributions only from the $W_4$ field into account, we get the mutual information
\be \label{in24}
I_n^{(2,4)}=I_n^{(2,4)tree}+I_n^{(2,4)1-loop}+I_n^{(2,4)2-loop}+I_n^{(2,4)3-loop}+\cdots,
\ee
where $I_n^{(2,4)tree}$, $I_n^{(2,4)2-loop}$, and $I_n^{(2,4)3-loop}$  being the same with $I_n^{(2)tree}$ (\ref{t0}), $I_n^{(2)2-loop}$ (\ref{t2}) and $I_n^{(2)3-loop}$ (\ref{t3}) respectively, but
\bea
&& I_n^{(2,4)1-loop}= \cdots +\frac{(n+1) x^8}{64023737057280000 n^{15}}\left(275290459983 n^{14}+3447342687823 n^{12} \rt. \nn\\
&& \phantom{I_n^{(2,4)1-loop}=}\lt.+24307301089071 n^{10}+168575843022951 n^8-40294949578091 n^6 \rt. \\
&& \phantom{I_n^{(2,4)1-loop}=}\lt.-11722057081371 n^4+1032493037037 n^2+2876199658597\right)+{O}(x^9). \nn
\eea
with $\cdots$ being the $x^4$, $x^5$, $x^6$ and $x^7$ parts of $I_n^{(2)1-loop}$ (\ref{t1}).
The 1-loop result is just the summation of $I_n^{(2)1-loop}$ (\ref{t1}) and $I_n^{spin-4,1-loop}$ (\ref{ins4}), and so is consistent with the bulk result.

\subsection{CFT with $W(2,3,4)$ symmetry}

Taking into account of both the contributions from $W_{3}$ and $W_{4}$ fields, we get the mutual information
\be \label{in234}
I_n^{(2,3,4)}=I_n^{(2,3,4)tree}+I_n^{(2,3,4)1-loop}+I_n^{(2,3,4)2-loop}+I_n^{(2,3,4)3-loop}+\cdots,
\ee
where $I_n^{(2,3,4)tree}$, $I_n^{(2,3,4)2-loop}$, and $I_n^{(2,3,4)3-loop}$  being the same with $I_n^{(2)tree}$ (\ref{t0}), $I_n^{(2,3)2-loop}$ (\ref{w2}), and $I_n^{(2)3-loop}$ (\ref{t3}) being just the summation of $I_n^{(2)1-loop}$ (\ref{t1}), $I_{n,spin-3}^{1-loop}$ (\ref{ins3}) and $I_{n,spin-4}^{1-loop}$ (\ref{ins4}). This is consistent with the bulk result as well.

Note that to order $O(x^8)$ we have the relation
\be
I_n^{(2,3,4)}=I_n^{(2,3)}+I_n^{(2,4)}-I_n^{(2)},
\ee
but it is generally not true for higher order of $x$.

\section{Conclusion and discussion} \label{s4}

In this paper we investigated the R\'enyi entropy of two intervals on complex plane with small cross ratio $x$ for 2D CFT with $W$ symmetry. The results could be written as the expansion of small $x$ to order $x^8$, and so only the $W_3$ and $W_4$ fields, and correspondingly the spin-3 and spin-4 fields in the gravity side, may contribute. We firstly did the calculation in the gravity side using the method in \cite{Hartman:2013mia,Faulkner:2013yia,Barrella:2013wja} and got the classical and 1-loop parts of the holographic R\'enyi entropy. In the calculation we assumed that the classical configurations are not changed by the presence of higher spin fields. Then we calculated in the CFT side using the OPE of twist operators, and found exact match with the gravity results. Our investigation provides a nontrivial test of HS/CFT correspondence in AdS$_3$ beyond the genus-1 partition function. Together with the results in \cite{Barrella:2013wja,Chen:2013kpa}, our study strongly support the holographic computation of R\'enyi entropy at classical and 1-loop level in both pure AdS$_3$ gravity and the higher spin AdS$_3$ gravities.

Our study shows that there are nonvanishing 2-loop and 3-loop corrections to the R\'enyi entropies $S_n, n>3$. This is related to the fact that the partition function on a higher genus Riemann surface is not 1-loop exact in dual gravity. It is certainly interesting to check the correspondence beyond the 1-loop level. In particular, there is weak sign that the $m$-loop correction with $m\geq 3$ is vanishing for $S_n,  n=2,3,\cdots,m$. It would be nice to check if this is true in CFT and gravity.

In this work we have considered a general CFT with $W$ symmetry. For a concrete CFT, for example the minimal model that is dual to higher spin gravitational theory with scalars in AdS$_3$ \cite{Gaberdiel:2010ar,Gaberdiel:2011zw,Gaberdiel:2012uj}, we have to consider the contributions from the scalars in the theory. In this so-called Gaberdiel-Gopakumar duality, there are many light states in the CFT which could contribute significantly to the R\'enyi entropies. However, it is not clear what kind of objects in the bulk they correspond to and if such objects contribute to the bulk computation.  We leave this interesting issue to future work.

Recently there has been calculation of R\'enyi entropy for logarithmic CFT in\cite{Chen:2014kja}. The computations in this work and \cite{Chen:2013kpa,Chen:2014kja} are straightforward but quite tedious, and also they are only limited to short interval expansion. It would be nice to find better ways to simplify the computations in order to extend the discussions to higher levels. General calculations beyond short interval expansion would also be desirable.

Another interesting question is about the relation between different R\'enyi entropies $S_n$.  As it is well-known that for two intervals on complex plane case $S_2$ is the genus-1 partition function which could be computed in CFT in other ways\cite{Headrick:2010zt}, and $S_n$ is expected to be the genus-$(n-1)$ partition function of the CFT. From the works \cite{Chen:2013kpa,Chen:2014kja}, once the genus-1 partition function is in match with the 1-loop bulk partition function, so do the higher R\'enyi entropies $S_n(n>2)$ match with the bulk partition function, at least to 1-loop. It would be great to see if or not this is true in general.

%In this paper we have calculated to three loops in the CFT side, but in the gravity side we can only get the tree and 1-loop parts. It would be great if there would be higher loops calculation in the gravity side to check the matches between the holographic R\'enyi entropy and the R\'enyi entropy.

In this work, we focused on the entanglement entropy in the ground state of CFT, without turning on the chemical potential for $W$ charges. In \cite{Ammon:2013hba,deBoer:2013vca}, the higher spin entanglement entropy has been discussed from gravity side. It would be interesting to investigate the higher spin R\'enyi entropies holographically or study them in CFT.

%\vspace*{1cm}
\noindent {\large{\bf Acknowledgments}} \\~\\
We thank Matthew Headrick for his Mathematica code Virasoro.nb that could be downloaded at his personal homepage \url{http://people.brandeis.edu/~headrick/Mathematica/index.html}.
BC would like to thank NCTS, Taiwan for hospitality during the final stage of this work.
We thank the anonymous referee of the paper for the valuable comments and suggestions.
The work was in part supported by NSFC Grant No.~11275010, No.~11335012 and No.~11325522.
JJZ was also in part supported by the Scholarship Award for Excellent Doctoral Student granted by the Ministry of Education of China.
\vspace*{1cm}

\begin{appendix}

\section{Counting of quasiprimary operators} \label{sa}

In this appendix we count the quasiprimary operators that are needed in the calculation, and get their normalization constant $\a_K$ defined in (\ref{e1}).
At the first step,  we work out the holomorphic quasiprimary operators constructed by the operators in the Virasoro Verma module to level eight for an ordinary CFT and then for the $CFT^n$. Next we find the additional holomorphic quasiprimary operators in the presence of $W$ fields. %The method of calculating the normalization coefficients of the operators $\a_K$ could be found in \cite{Chen:2013kpa}, and here we only give the results.

For the original CFT before taking the replica, the partition function of the unit operator 1 is
\be
\tr x^{L_0}=\prod_{m=2}^\inf \f{1}{1-x^m}=1+x^2+x^3+2x^4+2x^5+4x^6+4x^7+7x^8+ {O}(x^9),
\ee
where $\tr$ counts the number of holomorphic operators. So the number of linearly  independent holomorphic quasiprimary operators $\#$ at each level $L_0$ is
\begin{center} \begin{tabular}{c|c|c|c|c|c|c|c|c|c}
  $L_0$ & 0 & 2 & 3 & 4 & 5 & 6 & 7 & 8 & $\cdots$ \\\hline
  $\#$  & 1 & 1 & 0 & 1 & 0 & 2 & 0 & 3 & $\cdots$
\end{tabular} \end{center}
With some efforts, we could get the quasiprimary operators listed as follows.
\begin{itemize}
  \item At level 0, it is of course the identity $1$ with $\a_1=1$.
  \item At level 2, it is $T$ with $\a_T=\f{c}{2}$.
  \item At level 4, it is $\mc A=(TT)-\f{3}{10}\p^2T$ with $\a_{\mc A}=\f{c(5c+22)}{10}$.
  \item At level 6, they are
  \bea
  && \mc B=(\p T\p T)-\f{4}{5}(T\p^2T)+\f{23}{210}\p^4T,  \nn\\
  && \mc D=\mc C+\f{93}{70c+29} \mc B
  \eea
  with
  \bea
  \mc C=(T(TT))-\f{9}{10}(T\p^2 T)+\f{4}{35}\p^4 T.
  \eea
  Their normalization constants are respectively
  \bea
  && \a_{\mc B}=\frac{36c (70 c+29)}{175} , \nn\\
  && \a_{\mc D}=\frac{3 c (2 c-1) (5 c+22) (7 c+68)}{4 (70 c+29)}.
  \eea
  \item At level 8, they are
  \bea
  && \mc E= (\p^2 T\p^2 T)-\f{10}{9}(\p T \p^3T)+\f{10}{63}(T\p^4T)-\f{13}{2268}\p^6 T, \nn\\
  && \mc H=\mc F+ \frac{9 (140 c+83)}{50 (105 c+11)}\mc E, \nn\\
  && \mc I=\mc G+ \frac{81 (35 c-51)}{100 (105 c+11)}\mc E + \frac{12(465 c-127)}{5 c (210 c+661)-251}\mc H,
  \eea
  with
  \bea
  && \mc F=(T(\p T\p T))-\f{4}{5}(T(T\p^2 T))-\f{1}{5}(\p T \p^3T)+\f{47}{210}(T\p^4T)-\f{29}{2520}\p^6 T,   \nn\\
  && \mc G=(T(T(TT)))-\f{9}{5}(T(T\p^2 T))+\f{3}{10}(\p T \p^3T)+\f{29}{70}(T\p^4T)-\f{41}{1680}\p^6 T.
  \eea
  The corresponding normalization constants are respectively
  \bea
  && \a_{\mc E}=\frac{22880 c (105 c+11)}{1323}, \nn\\
  && \a_{\mc H}=\frac{26 c (5 c+22) (5 c (210 c+661)-251)}{125 (105 c+11)},\nn\\
  && \a_{\mc I}=\frac{3 c (2 c-1) (3 c+46) (5 c+3) (5 c+22) (7 c+68)}{2 (5 c (210 c+661)-251)}.
  \eea
\end{itemize}

For the $CFT^n$, we consider the partition function
\bea
&& \tr x^{L_0}=\prod_{m=2}^\inf \f{1}{(1-x^m)^n}=1+n x^2+n x^3+\frac{n (n+3)}{2} x^4+n (n+1) x^5+\frac{ n (n+1) (n+11)}{6} x^6 \nn\\
&& \phantom{\tr x^{L_0}=\prod_{m=2}^\inf \f{1}{(1-x^m)^n}=}
+\frac{ n (n^2+5n +2)}{2} x^7+\frac{ n (n+3) (n^2+27n+14)}{24} x^8+{O}(x^9),
\eea
where $\tr$ counts the holomorphic operators of the $CFT^n$. The number of linear independent holomorphic quasiprimary operators \# at each level $L_0$ is
\begin{center} \begin{tabular}{c|c|c|c|c|c|c|c|c|c|c}
$L_0$ & 0 & 1 & 2 & 3 & 4 & 5 & 6 & 7 & 8 & $\cdots$ \\ \hline
\# & 1 & 0 & $n$ & 0 & $\f{n(n+1)}{2}$ & $\f{n(n-1)}{2}$ & $\f{n(n+1)(n+5)}{6}$ & $\f{n(n-1)(2n+5)}{6}$ & $\f{n(n+1)(n^2+17n+18)}{24}$ & $\cdots$ \\
\end{tabular} \end{center}
The quasiprimary operators are listed in the following table, in which $j_i$'s are integer, taking values between $0$ and $n-1$.
\begin{center} \begin{tabular}{|c|c|c|c|}\hline
    $L_0 $ & quasiprimary operators & degeneracies & \# \\\hline
    0 & 1 & 1 & 1\\ \hline

    2 & $T_j$ & $n$ & $n$ \\\hline
    \multirow{2}*{4} & $\mc A_j$ & $n$ & \multirow{2}*{$\f{n(n+1)}{2}$} \\ \cline{2-3}
       &$T_{j_1}T_{j_2}$ with $j_1 < j_2$ & $\f{n(n-1)}{2}$ & \\\hline

    5 & $\mc J_{j_1 j_2}$ with $j_1 < j_2$ & $\f{n(n-1)}{2}$ & $\f{n(n-1)}{2}$ \\ \hline
      & $\mc B_j$ & $n$ & \\ \cline{2-3}
      & $\mc D_j$ & $n$ & \\ \cline{2-3}
    6 & $T_{j_1} \mc A_{j_2}$ with $j_1 \neq j_2$ & $n(n-1)$ &  $\f{n(n+1)(n+5)}{6}$ \\ \cline{2-3}
      & $\mc K_{j_1j_2}$ with $j_1 < j_2$ & $\f{n(n-1)}{2}$ &     \\ \cline{2-3}
      & $T_{j_1}T_{j_2}T_{j_3}$ with $j_1 < j_2<j_3$ & $\f{n(n-1)(n-2)}{6}$ & \\ \hline

      & $\mc L_{j_1j_2}$ with $j_1\neq j_2$ & $n(n-1)$ & \\  \cline{2-3}
    \multirow{2}*{7} & $\mc M_{j_1j_2}$ with $j_1 < j_2$ & $\f{n(n-1)}{2}$ & \multirow{2}*{$\f{n(n-1)(2n+5)}{6}$} \\  \cline{2-3}
      & $T_{j_1}\mc J_{j_2j_3}$ with $j_1<j_2<j_3$ & \multirow{2}*{$\f{n(n-1)(n-2)}{3}$} & \\  \cline{2-2}
      & $\mc N_{j_1j_2j_3}$ with $j_1<j_2<j_3$ & & \\  \hline

      & $\mc E_j$ & $n$ & \\ \cline{2-3}
      & $\mc H_j$ & $n$ & \\ \cline{2-3}
      & $\mc I_j$ & $n$ & \\ \cline{2-3}
      & $T_{j_1} \mc B_{j_2}$ with $j_1 \neq j_2$ & $n(n-1)$ &  \\ \cline{2-3}
      & $T_{j_1} \mc D_{j_2}$ with $j_1 \neq j_2$ & $n(n-1)$ &  \\ \cline{2-3}
      & $\mc A_{j_1} \mc A_{j_2}$ with $j_1 < j_2$ & $\f{n(n-1)}{2}$ &  \\ \cline{2-3}
   8  & $\mc O_{j_1j_2}$ with $j_1 \neq j_2$ & $n(n-1)$ &  $\f{n(n+1)(n^2+17n+18)}{24}$  \\ \cline{2-3}
      & $\mc P_{j_1j_2}$ with $j_1 < j_2$ & $\f{n(n-1)}{2}$ &     \\ \cline{2-3}
      & $T_{j_1}T_{j_2}\mc A_{j_3}$ with $j_1 < j_2$, $j_1\neq j_3$ and $j_2 \neq j_3$ & $\f{n(n-1)(n-2)}{2}$ &     \\ \cline{2-3}
      & $T_{j_1}\mc K_{j_2j_3}$ with $j_1 < j_2<j_3$ &  &     \\ \cline{2-2}
      & $\mc Q_{j_1j_2j_3}$ with $j_1 < j_2<j_3$ & $\f{n(n-1)(n-2)}{2}$ &     \\ \cline{2-2}
      & $\mc R_{j_1j_2j_3}$ with $j_1 < j_2<j_3$ &  &     \\ \cline{2-3}
      & $T_{j_1}T_{j_2}T_{j_3}T_{j_4}$ with $j_1 < j_2<j_3<j_4$ & $\f{n(n-1)(n-2)(n-3)}{24}$ & \\ \hline

    $\cdots$ & $\cdots$ & $\cdots$ & $\cdots$\\
    \hline
\end{tabular} \end{center}
In the above table, there are several new quasiprimary operators, which could not be written as the product of the quasiprimary operators in different replicas. At level five and six, they are
\bea
&& \mc J_{j_1j_2}=T_{j_1}i\p T_{j_2}-i\p T_{j_1}T_{j_2},  \nn\\
&& \mc K_{j_1j_2}=\p T_{j_1} \p T_{j_2}-\f{2}{5} \lt( T_{j_1}\p^2 T_{j_2}+\p^2 T_{j_1}T_{j_2} \rt).
\eea
At level seven they are
\bea
&& \mc L_{j_1j_2}=T_{j_1}i\p\mc A_{j_2}-2 i\p T_{j_1}\mc A_{j_2},  \nn\\
&& \mc M_{j_1j_2}=i\p T_{j_1}\p^2 T_{j_2}-\p^2 T_{j_1}i\p T_{j_2}+\f{2}{9} \lt( i\p^3T_{j_1}T_{j_2}-T_{j_1}i\p^3T_{j_2} \rt), \nn\\
&& \mc N_{j_1j_2j_3}=T_{j_1}\lt( T_{j_2}i\p T_{j_3} + i\p T_{j_2}T_{j_3} \rt)-2i\p T_{j_1}T_{j_2}T_{j_3}.
\eea
At level eight they are
\bea
&& \mc O_{j_1j_2}=\p T_{j_1}\p\mc A_{j_2}-\f{2}{9}T_{j_1}\p^2 \mc A_{j_2}-\f{4}{5}\p^2 T_{j_1}\mc A_{j_2}, \nn\\
&& \mc P_{j_1j_2}=\p^2 T_{j_1}\p^2 T_{j_2}-\f{5}{9} \lt( \p T_{j_1}\p^3 T_{j_2} + \p^3 T_{j_1}\p T_{j_2} \rt)
                                          +\f{5}{63} \lt( T_{j_1}\p^4 T_{j_2} +\p^4 T_{j_1}T_{j_2} \rt), \nn\\
&& \mc Q_{j_1j_2j_3}=T_{j_2}K_{j_3j_1}-\f{2}{9}T_{j_1}K_{j_2j_3},  \nn\\
&& \mc R_{j_1j_2j_3}=T_{j_3}K_{j_1j_2}-\f{2}{11}\lt( T_{j_1}K_{j_2j_3}+T_{j_2}K_{j_3j_1}  \rt).
\eea

The normalization factors of these orthogonalized quasiprimary operators are respectively
\bea
&& \a_{TT}=\f{c^2}{4}, ~~~
\a_\mJ=2c^2, ~~~
\a_{T \mA}=\f{c^2(5c+22)}{20}, ~~~
\a_\mK=\f{36c^2}{5},\nn\\
&& \a_{TTT}=\f{c^3}{8}, ~~~
\a_\mL=\f{6c^2(5c+22)}{5}, ~~~
\a_\mM=\f{880c^2}{9},\nn\\
&& \a_{T\mJ}=c^3, ~~~
\a_\mN=3c^3, ~~~
\a_{T\mc B}=\frac{18c^2 (70 c+29)}{175},\nn\\
&& a_{T\mc D}=\frac{3 c^2 (2 c-1) (5 c+22) (7 c+68)}{8 (70 c+29)}, \\
&& \a_{\mA\mA}=\f{c^2(5c+22)^2}{100}, ~~~
\a_\mO=\f{728c^2(5c+22)}{225}, \nn\\
&& \a_\mP=\f{57200c^2}{63}, ~~~
\a_{TT\mA}=\f{c^3(5c+22)}{40}, ~~~ \a_{T\mK}=\f{18c^3}{5}, \nn\\
&& \a_{\mQ}=\f{154c^3}{45}, ~~~
\a_{\mc R}=\f{182c^3}{55}, ~~~
\a_{TTTT}=\f{c^4}{16}. \nn
\eea

For a CFT with $W$ symmetry, we should consider additional contributions, and the partition function is
\be
\tr x^{L_0}=\prod_{s}\prod_{m=s}^\inf\frac{1}{1-x^m},
\ee
where the $\tr$ counts the number of the holomorphic operators again, and $s$ denotes all the possible spins in the theory.
If the theory has $W(2,3)$ symmetry, it is
\be
\tr x^{L_0}=\frac{1}{1-x^2}\prod_{m=3}^\inf\frac{1}{\lt(1-x^m\rt)^2}
           =1+x^2+2 x^3+3 x^4+4 x^5+8 x^6+10 x^7+17 x^8+O(x^9),
\ee
and for $CFT^n$ it would be
\bea
&& \tr x^{L_0}=\frac{1}{(1-x^2)^n}\prod_{m=3}^\inf\frac{1}{\lt(1-x^m\rt)^{2n}} \nn\\
&& \phantom{\tr x^{L_0}}
           =1+n x^2+2 n x^3+\frac{n (n+5)}{2} x^4+2 n (n+1) x^5+\frac{n (n^2+27n +20)}{6} x^6\nn\\
&& \phantom{\tr x^{L_0}=}
           +n (n^2+7n+2) x^7+\frac{n (n+3) (n^2+75n+26)}{24} x^8+O(x^9).
\eea
The numbers of independent quasiprimary operators for the CFT and the $CFT^n$ are listed as follows.
\begin{center} \begin{tabular}{c|c|c|c|c|c|c|c|c|c|c}
$L_0$ & 0 & 1 & 2 & 3 & 4 & 5 & 6 & 7 & 8 & $\cdots$ \\ \hline
CFT   & 1 & 0 & 1 & 1 & 1 & 1 & 4 & 2 & 7 & $\cdots$ \\ \hline
$CFT^n $ & 1 & 0 & $n$ & $n$ & $\f{n(n+1)}{2}$ & $\f{n(3n-1)}{2}$ & $\f{n(n^2+15n+8)}{6}$ & $\f{n (5n^2+15n-8)}{6}$ & $\f{n(n+1)(n^2+53n+30)}{24}$ & $\cdots$ \\
\end{tabular} \end{center}
If the theory has $W(2,4)$ symmetry, it is
\be
\tr x^{L_0}=\frac{1}{(1-x^2)(1-x^3)}\prod_{m=4}^\inf\frac{1}{\lt(1-x^m\rt)^2}
           =1+x^2+x^3+3 x^4+3 x^5+6 x^6+7 x^7+13 x^8+O(x^9),
\ee
and for $CFT^n$ it would be
\bea
&& \tr x^{L_0}=\frac{1}{(1-x^2)^n(1-x^3)^n}\prod_{m=4}^\inf\frac{1}{\lt(1-x^m\rt)^{2n}} \nn\\
&& \phantom{\tr x^{L_0}}
           =1+n x^2+n x^3+\frac{n (n+5)}{2} x^4+n (n+2) x^5+\frac{ n (n+1) (n+17)}{6} x^6\nn\\
&& \phantom{\tr x^{L_0}=}
           +\frac{n (n^2+9n+4)}{2} x^7+\frac{n ( n^3+42n^2+191n+78)}{24} x^8+O(x^9).
\eea
Then we can get the numbers of independent quasiprimary operators for the CFT and the $CFT^n$ listed as follows.
\begin{center} \begin{tabular}{c|c|c|c|c|c|c|c|c|c|c}
$L_0$ & 0 & 1 & 2 & 3 & 4 & 5 & 6 & 7 & 8 & $\cdots$ \\ \hline
CFT   & 1 & 0 & 1 & 0 & 2 & 0 & 3 & 1 & 6 & $\cdots$ \\ \hline
$CFT^n $ & 1 & 0 & $n$ & 0 & $\f{n(n+3)}{2}$ & $\f{n(n-1)}{2}$ & $\frac{n(n^2+12n+5)}{6}$ & $\frac{n (n+5) (2 n-1)}{6}$ & $\frac{n(n^3+30n^2+83n+30)}{24}$ & $\cdots$ \\
\end{tabular} \end{center}
If the theory has $W(2,3,4)$ symmetry, it is
\be
\tr x^{L_0}=\frac{1}{(1-x^2)(1-x^3)^2}\prod_{m=4}^\inf\frac{1}{\lt(1-x^m\rt)^3}
           =1+x^2+2 x^3+4 x^4+5 x^5+10 x^6+14 x^7+25 x^8+O(x^9),
\ee
and for $CFT^n$ it would be
\bea
&& \tr x^{L_0}=\frac{1}{(1-x^2)^n(1-x^3)^{2n}}\prod_{m=4}^\inf\frac{1}{\lt(1-x^m\rt)^{3n}} \nn\\
&& \phantom{\tr x^{L_0}}
           =1+n x^2+2 n x^3+\frac{n (n+7)}{2}x^4+n (2 n+3) x^5+\frac{n \left(n^2+33n+26\right)}{6}x^6\nn\\
&& \phantom{\tr x^{L_0}=}
            +n \left(n^2+10n+3\right) x^7+\frac{n \left(n^3+90n^2+395n+114\right)}{24}x^8+O(x^9).
\eea
Then the numbers of independent quasiprimary operators for the CFT and the $CFT^n$ are listed as follows.
\begin{center} \begin{tabular}{c|c|c|c|c|c|c|c|c|c|c}
$L_0$ & 0 & 1 & 2 & 3 & 4 & 5 & 6 & 7 & 8 & $\cdots$ \\ \hline
CFT   & 1 & 0 & 1 & 1 & 2 & 1 & 5 & 4 & 11 & $\cdots$ \\ \hline
$CFT^n $ & 1 & 0 & $n$ & $n$ & $\f{n(n+3)}{2}$ & $\f{n(3n-1)}{2}$ & $\frac{n(n^2+21n+8)}{6}$ & $\frac{n (5n^2+27n-8)}{6}$ & $\frac{n(n^3+66n^2+155n+42)}{24}$ & $\cdots$ \\
\end{tabular} \end{center}
There are large numbers of the quasiprimary operators from $W$ fields. However, most of them  have vanishing coefficients $d_K$ so that they do not contribute to the OPE of the twist operators.

For the CFT with $W(2,3)$ ($W_3$) symmetry, we denote the extra holomorphic and antiholomorphic primary operator as $W(z)=W_3(z)$ and $\bar W(\bar z)=\bar W_3(\bar z)$ which have conformal weight $(3,0)$ and $(0,3)$ respectively. In this case to level eight for $CFT^n$, the additional holomorphic quasiprimary operators with nonvanishing coefficients $d_K$ are listed as below.
\begin{center} \begin{tabular}{|c|c|c|}\hline
$L_0 $ & quasiprimary operators & degeneracies  \\\hline
6 & $W_{j_1}W_{j_2}$ with $j_1 < j_2$ & $\f{n(n-1)}{2}$ \\ \hline

7 & $\mU_{j_1 j_2}$ with $j_1 < j_2$ & $\f{n(n-1)}{2}$  \\\hline

  & $W_{j_1}\mS_{j_2}$ with $j_1 \neq j_2$ & $n(n-1)$  \\ \cline{2-3}
8 & $\mV_{j_1 j_2}$ with $j_1 < j_2$ & $\f{n(n-1)}{2}$  \\\cline{2-3}
  & $T_{j_1}W_{j_2}W_{j_3}$ with $j_1 \neq j_2$, $j_1 \neq j_3$ and $j_2<j_3$  & $\f{n(n-1)(n-2)}{2}$ \\ \hline

$\cdots$ & $\cdots$ & $\cdots$\\
\hline
\end{tabular} \end{center}
Here we have
\bea \label{e3}
&& \mS=(TW)-\f{3}{14}\p^2 W, \nn\\
&& \mU_{j_1j_2}=W_{j_1}i\p W_{j_2}-i\p W_{j_1}W_{j_2},  \nn\\
&& \mV_{j_1j_2}=\p W_{j_1} \p W_{j_2}-\f{2}{7} \lt( W_{j_1}\p^2 W_{j_2}+\p^2 W_{j_1}W_{j_2} \rt).
\eea
As we mentioned there are actually some other quasiprimary operators, for examples $W_{j}$ at level three and  $T_{j_1}W_{j_2}$ with $j_1 \neq j_2$ at level five, but all of them have vanishing coefficients $d_K$, and so do not contribute to the OPE of twist operators.  The normalization factor of $W(z)$ is
$\a_W=\f{c}{3}$ such that the normalization factors of the above quasiprimary operators are respectively
\bea
&&  \a_\mS=\f{c(7c+114)}{42}, ~~~ \a_{WW}=\f{c^2}{9}, ~~~ \a_\mU=\f{4c^2}{3}, \nn\\
&& \a_{W\mS}=\f{c^2(7c+114)}{126}, ~~~ \a_\mV=\f{52c^2}{7}, ~~~ \a_{TWW}=\f{c^3}{18}.
\eea
Obviously, if we only consider the operators up to level 6, the quasiprimary operators from $W$ fields are somehow trivial, without mixing with the stress tensor.

For the CFT with $W(2,4)$ symmetry, we denote the extra holomorphic and antiholomorphic primary operators as $W_4(z)$ and $\bar W_4(\bar z)$, which has conformal weight $(4,0)$ and $(0,4)$ respectively. In this case to level eight for $CFT^n$, the additional holomorphic quasiprimary operators are $W_{4j_1}W_{4j_2}$ with $j_1<j_2$. We choose the normalization such that
\be
\a_{W_4}=\a_{W_4W_4}=1.
\ee

For the CFT with $W(2,3,4)$ ($W_4$) symmetry, the extra holomorphic quasiprimary operators to level eight consist of the ones from $W_3$ field and the ones from $W_4$ field listed above. One may wonder if there exist the quasiprimary operators that are the mixtures of the stress tensor, $W_3$ and $W_4$ fields. Such kind of operators do exist, for example $W_{3j_1}W_{4j_2}$ with $j_1 \neq j_2$ at level seven, but they having vanishing coefficients $d_K$ and do not contribute.

\section{The coefficients $d_K$} \label{sb}

The calculation of the OPE coefficients $d_K$ for the quasiprimary operators  is straightforward but tedious\cite{Calabrese:2010he,Chen:2013kpa}. For brevity, we only outline the process and then give the results. Firstly we need to determine how the holomorphic operators $T(z)$, $\mA(z)$, $\mB(z)$, $\mD(z)$, $\mE(z)$, $\mH(z)$ and $\mI(z)$ transform under the coordinate transformation $z \to f(z)$ with $f(z)$ being a general holomorphic function of $z$. We denote the Schwarz derivative as
\be
s(z) \equiv \f{f'''(z)}{f'(z)}-\f{3}{2} \lt( \f{f''(z)}{f'(z)} \rt)^2,
\ee
and for simplicity we use the shorthands
\bea
&& f=f(z), ~~~ f'=f'(z), ~~~ f''=f''(z), ~~~ \cdots  \nn\\
&& s=s(z), ~~~ s'=s'(z), ~~~ s''=s''(z), ~~~ \cdots.
\eea
The transformations of $T$ and $\mA$ are
\bea
&& T(z)=f'^2 T(f)+\f{c}{12}s, \nn\\
&& \mA(z)=f'^4\mA(f)+\f{5c+22}{30}s \lt( f'^2 T(f)+\f{c}{24}s \rt).
\eea
The transformation of one single operator include the homogeneous terms, which are proportional to nonidentity quasiprimary operators or derivatives thereof, and the inhomogeneous terms which are proportional to the identity operator. For examples, the transformation of $T(z)$ includes one homogeneous term $f'^2 T(f)$ and one inhomogeneous term $\f{c}{12}s$, while the transformation of $\mA(z)$ includes two homogeneous terms $f'^4\mA(f)+\f{5c+22}{30} f'^2 s T(f)$ and one inhomogeneous term $\f{c(5c+22)s^2}{720}$. The transformation of $\mB(z)$ is complicated
\bea
&& \mB(z)=f'^6\mB(f)-\f{8}{5}f'^4s\mA(f) \nn\\
&& \phantom{\mB(z)=}
        -\f{1}{1050}\lt( 28(5c+22)f'^2s^2+(70c+29)(f'^2s''-5f'f''s'+5f''^2s) \rt)T(f) \nn\\
&& \phantom{\mB(z)=}
   +\f{70c+29}{420}f'^2(f's'-2f''s)\p T(f)-\f{70c+29}{1050}f'^4s\p^2T(f) \nn\\
&& \phantom{\mB(z)=}
   -\f{c}{50400}\lt( 744s^3+ (70c+29)(4ss''-5s'^2) \rt),
\eea
and the transformation of $\mD(z)$ is relatively easy
\be
\mD(z)=f'^6\mD(f)+\f{(2c-1)(7c+68)}{70c+29}s \lt( \f{5}{4} f'^4\mA(f)+\f{5c+22}{48}s \lt( f'^2T(f)+\f{c}{36}s \rt) \rt).
\ee
In the following discussion we only need the inhomogeneous terms of the transformations of $\mE(z)$, $\mH(z)$ and $\mI(z)$, and explicitly we have
\bea
&& \mE(z)=f'^8 \mE(f)+\cdots \nn\\
&& \phantom{\mE(z)=}
+\frac{c}{952560} \left( 23452s^4+2255s(4ss''-5s'^2)+(105 c+11)(10ss^{(4)}-70s's'''+63s''^2)\right), \nn\\
&& \mH(z)=f'^8 \mH(f)+\cdots \nn\\
&& \phantom{\mH(z)=}
-\frac{c (5 c+22)}{1296000 (105 c+11)} s
\left( 104(465 c-127) s^3 +3 \lt(5 c (210 c+661)-251\rt) (4 s s''-5 s'^2) \right), \nn\\
&& \mI(z)=f'^8\mI(f)+\cdots
+\frac{c (2 c-1) (3 c+46) (5 c+3) (5 c+22) (7 c+68)}{20736 (5 c (210 c+661)-251)}s^4,
\eea
with the $\cdots$'s represent some homogeneous terms.

The coefficients $d_K$ for identity and quasiprmary operators  at only one replica could be calculated easily as
\bea
&&  d_1=1, ~~~ d_T=\frac{n^2-1}{12n^2}, ~~~ d_\mA=\frac{(n^2-1)^2}{288 n^4},
~~~ d_\mB=-\frac{(n^2-1)^2 \left(2 n^2(35 c+61)-93\right)}{10368 n^6(70 c+29)}, \nn\\
&& d_\mD=\frac{(n^2-1)^3}{10368 n^6},
~~~ d_\mE=\frac{(n^2-1)^2 \left(11340 n^4 c+11561 n^4-16236 n^2+5863\right)}{65894400 n^8(105 c+11)}, \\
&& d_\mH=-\frac{(n^2-1)^3 \left(3150 n^2 c^2 + \left(15960 n^2-6045\right)c-\left(2404 n^2-1651\right)\right)}{539136 n^8(5 c (210 c+661)-251)},
~~~ d_\mI=\frac{(n^2-1)^4}{497664 n^8}. \nn
\eea
For quasiprimary operators that can be factorized as two nonidentity quasiprimary operators at two different replicas, we need the orthogonal relation of the operators. And the final results are
\bea
&& d_{TT}^{j_1j_2}=\f{1}{8n^4c}\f{1}{s^4_{j_1j_2}}+\f{(n^2-1)^2}{144n^4},
~~~ d_{T\mA}^{j_1j_2}=\f{n^2-1}{96n^6c}\f{1}{s^4_{j_1j_2}}+\f{(n^2-1)^3}{3456n^6}, \nn\\
&& d_{T\mB}^{j_1j_2}=\f{5(n^2-1)}{6912n^8c}\f{1}{s_{j_1j_2}^6}
                     -\f{(n^2-1)\lt( 140(n^2+2)c+337n^2-163 \rt)}{6912n^8c(70c+29)}\f{1}{s_{j_1j_2}^4} \nn\\
&& \phantom{d_{T\mB}^{j_1j_2}=}
                     -\f{(n^2-1)^3\lt( 2n^2(35c+61)-93 \rt)}{124416n^8(70c+29)},  \\
&& d_{T\mD}^{j_1j_2}=\f{(n^2-1)^2}{2304n^8c}\f{1}{s_{j_1j_2}^4}+\f{(n^2-1)^4}{124416n^8},
~~~ d_{\mA\mA}^{j_1j_2}=\f{5}{128n^8c(5c+22)}\f{1}{s_{j_1j_2}^8}+\f{(n^2-1)^2}{1152n^8c}\f{1}{s_{j_1j_2}^4}+\f{(n^2-1)^4}{82944n^8},\nn
\eea
where we have defined $s_{j_1j_2}\equiv\sin\f{\pi(j_1-j_2)}{n}$. For quasiprimary operators that can be factorized as three and four nonidentity quasiprimary operators at  different replicas, we need the correlation functions of three and four quasiprimary operators. The correlation functions we need are
\bea
&& \lag T(f_1) T(f_{2}) T(f_{3}) \rag_C=\f{c}{f_{12}^2f_{23}^2f_{13}^2}, \nn\\
&& \lag T(f_1) T(f_{2}) \mA(f_{3}) \rag_C=\f{c(5c+22)}{10}\f{1}{f_{23}^4f_{13}^4}, \\
&& \lag T(f_1) T(f_{2}) T(f_{3}) T(f_{4}) \rag_C=c \lt( \f{1}{\lt( f_{12}f_{34}f_{13}f_{24} \rt)^2}
                                                       +\f{1}{\lt( f_{13}f_{24}f_{14}f_{23} \rt)^2}
                                                       +\f{1}{\lt( f_{12}f_{34}f_{14}f_{23} \rt)^2} \rt) \nn\\
&& \phantom{\lag T(f_1) T(f_{2}) T(f_{3}) T(f_{4}) \rag_C=}
                                       +\f{c^2}{4} \lt( \f{1}{\lt( f_{12}f_{34}\rt)^4}
                                                       +\f{1}{\lt( f_{13}f_{24}\rt)^4}
                                                       +\f{1}{\lt( f_{14}f_{23} \rt)^4} \rt), \nn
\eea
where $f_{ij}\equiv f_i-f_j$. And then we get
\bea
&& d_{TTT}^{j_1j_2j_3}=-\f{1}{8n^6c^2}\f{1}{s^2_{j_1j_2}s^2_{j_2j_3}s^2_{j_3j_1}}
                       +\f{n^2-1}{96n^6c} \lt( \f{1}{s^4_{j_1j_2}}+\f{1}{s^4_{j_2j_3}}+\f{1}{s^4_{j_3j_1}} \rt)
                       +\f{(n^2-1)^3}{1728n^6}, \nn\\
&& d_{TT\mA}^{j_1j_2j_3}=\f{1}{64n^8c^2}\f{1}{(s_{j_1j_3}s_{j_2j_3})^4}
                         -\f{n^2-1}{96n^8c^2}\f{1}{(s_{j_1j_2}s_{j_1j_3}s_{j_2j_3})^2} \nn\\
&& \phantom{d_{TT\mA}^{j_1j_2j_3}=}
                         +\f{(n^2-1)^2}{2304n^8c} \lt( \f{1}{s_{j_1j_2}^4}+\f{2}{s_{j_1j_3}^4}+\f{2}{s_{j_2j_3}^4} \rt)
                         +\f{(n^2-1)^4}{41472n^8} ,  \\
&& d_{TTTT}^{j_1j_2j_3j_4}= \f{1}{16n^8c^3} g_\a^{j_1j_2j_3j_4}
                            +\f{1}{64n^8c^2} g_\b^{j_1j_2j_3j_4}
                            -\f{n^2-1}{96n^8c^2} g_\g^{j_1j_2j_3j_4}
                            +\f{(n^2-1)^2}{1152n^8c} g_\d^{j_1j_2j_3j_4}
                            +\f{(n^2-1)^4}{20736n^8},  \nn
\eea
where
\bea \label{gabgd}
&& g_\a^{j_1j_2j_3j_4}=\f{1}{\lt( s_{j_1j_2}s_{j_3j_4}s_{j_1j_3}s_{j_2j_4} \rt)^2}
                       +\f{1}{\lt( s_{j_1j_3}s_{j_2j_4}s_{j_1j_4}s_{j_2j_3} \rt)^2}
                       +\f{1}{\lt( s_{j_1j_2}s_{j_3j_4}s_{j_1j_4}s_{j_2j_3} \rt)^2}  , \nn\\
&& g_\b^{j_1j_2j_3j_4}=\f{1}{\lt( s_{j_1j_2}s_{j_3j_4} \rt)^4}
                      +\f{1}{\lt( s_{j_1j_3}s_{j_2j_4}\rt)^4}
                      +\f{1}{\lt( s_{j_1j_4}s_{j_2j_3} \rt)^4}, \nn\\
&& g_\g^{j_1j_2j_3j_4}=\f{1}{\lt( s_{j_1j_2}s_{j_2j_3}s_{j_3j_1}\rt)^2}
                        +\f{1}{\lt( s_{j_1j_2}s_{j_2j_4}s_{j_4j_1}\rt)^2}
                        +\f{1}{\lt( s_{j_1j_3}s_{j_3j_4}s_{j_4j_1}\rt)^2}
                        +\f{1}{\lt( s_{j_2j_3}s_{j_3j_4}s_{j_4j_2}\rt)^2}, \nn\\
&& g_\d^{j_1j_2j_3j_4}=\f{1}{s_{j_1j_2}^4}
                       +\f{1}{s_{j_1j_3}^4}
                       +\f{1}{s_{j_1j_4}^4}
                       +\f{1}{s_{j_2j_3}^4}
                       +\f{1}{s_{j_2j_4}^4}
                       +\f{1}{s_{j_3j_4}^4}.
\eea
There are other operators that cannot be factorized, and the coefficients $d_K$ for these operators are
\bea
&& d_{\mJ}^{j_1j_2}=\f{1}{16n^5 c}\f{c_{j_1j_2}}{s^5_{j_1j_2}},
~~~ d_{\mK}^{j_1j_2}=\f{5}{128n^6c}\f{1}{s^6_{j_1j_2}}-\f{n^2+9}{288n^6c}\f{1}{s^4_{j_1j_2}}-\f{(n^2-1)^2}{5184n^4}, \nn\\
&& d_{\mL}^{j_1j_2}=\f{n^2-1}{384n^7 c}\f{c_{j_1j_2}}{s^5_{j_1j_2}},
~~~ d_{\mM}^{j_1j_2}=\f{3}{256n^7c}\f{c_{j_1j_2}}{s^7_{j_1j_2}}-\f{4n^2+11}{1760n^7c}\f{c_{j_1j_2}}{s^5_{j_1j_2}} ,\nn\\
&& d_{T\mJ}^{j_1j_2j_3}= -\f{1}{64 n^7 c^2}\f{2c_{j_2j_3}s_{j_1j_2}s_{j_1j_3}-s_{j_2j_3}^2}{(s_{j_1j_2}s_{j_1j_3}s_{j_2j_3})^3}
          -\f{n^2-1}{384n^7c} \lt( \f{c_{j_1j_2}}{s_{j_1j_2}^5}-\f{c_{j_1j_3}}{s_{j_1j_3}^5}-\f{2c_{j_2j_3}}{s_{j_2j_3}^5} \rt), \nn\\
&& d_\mN^{j_1j_2j_3}=\f{1}{3} \lt( d_{T\mJ}^{j_1j_2j_3}+2d_{T\mJ}^{j_3j_1j_2} \rt),
~~~ d_\mO^{j_1j_2}= \f{n^2-1}{39936n^8c}\f{65-4(2n^2+13)s_{j_1j_2}^2}{s_{j_1j_2}^6}-\f{(n^2-1)^3}{179712n^6}, \nn\\
&& d_{\mP}^{j_1j_2}=\f{1}{7321600n^8c}\f{30030-220(41n^2+169)s_{j_1j_2}^2+16(27n^4+451n^2+572)s_{j_1j_2}^4}{s_{j_1j_2}^8}
                    +\f{3(n^2-1)^2}{915200n^4}, \nn\\
&& d_{T\mK}^{j_1j_2j_3}=-\f{1}{2304n^8c^2}
                        \f{14s_{j_2j_3}^4+27s_{j_1j_2}^2s_{j_1j_3}^2-8s_{j_2j_3}^2(s_{j_1j_2}^2+s_{j_1j_3}^2)-8(n^2+2)(s_{j_1j_2}s_{j_1j_3}s_{j_2j_3})^2}
                        {(s_{j_1j_2}s_{j_1j_3}s_{j_2j_3})^4} \nn\\
&& \phantom{d_{T\mK}^{j_1j_2j_3}=}
+\f{n^2-1}{6912n^8c}\lt( \f{5-2(n^2+2)s_{j_1j_2}^2}{s_{j_1j_2}^6}
                         +\f{5-2(n^2+2)s_{j_1j_3}^3}{s_{j_1j_3}^6}
                         +\f{45-4(n^2+9)s_{j_2j_3}^2}{2s_{j_2j_3}^6}   \rt) \nn\\
&& \phantom{d_{T\mK}^{j_1j_2j_3}=}
-\f{(n^2-1)^3}{62208n^6},  \nn\\
&& d_\mQ^{j_1j_2j_3}=\f{81}{77}\lt( d_{T\mK}^{j_2j_3j_1}-\f{2}{9}d_{T\mK}^{j_1j_2j_3} \rt),
~~~ d_\mR^{j_1j_2j_3}=\f{99}{91}\lt( d_{T\mK}^{j_3j_1j_2}-\f{2}{11} \lt( d_{T\mK}^{j_1j_2j_3}+d_{T\mK}^{j_2j_3j_1} \rt) \rt).
\eea
Here $c_{j_1j_2}\equiv\cos\f{\pi(j_1-j_2)}{n}$.

In the presence of $W_3$ symmetry, the additional quasiprimary operators listed in the table above (\ref{e3}) have the coefficients
\bea
&& d_{WW}^{j_1j_2}=-\f{3}{(2n)^6c}\f{1}{s_{j_1j_2}^6},
~~~ d_\mU^{j_1j_2}=-\f{3}{(2n)^7c}\f{c_{j_1j_2}}{s_{j_1j_2}^7},
~~~ d_{W\mS}^{j_1j_2}=-\f{n^2-1}{(2n)^8c}\f{1}{s_{j_1j_2}^6},   \\
&& d_\mV^{j_1j_2}=\f{1}{26(2n)^8c} \lt( \f{6(n^2+13)}{s_{j_1j_2}^6}-\f{91}{s_{j_1j_2}^8}  \rt),
~~~ d_{TWW}^{j_1j_2j_3}=\f{18}{(2n)^8c^2}\f{1}{s_{j_1j_2}^2s_{j_2j_3}^4s_{j_3j_1}^2}
                      -\f{n^2-1}{(2n)^8c}\f{1}{s_{j_2j_3}^6}. \nn
\eea
In the calculation we need to use the three-point function
\be
\lag T(f_1)W(f_2)W(f_3) \rag_C=\f{c}{f_{12}^2 f_{13}^2 f_{23}^4}.
\ee
In the presence of $W_4$ symmetry we have
\be
d_{W_4W_4}^{j_1j_2}=\f{1}{(2n)^8}\f{1}{s_{j_1j_2}^8}.
\ee

\section{Some details in Subsection \ref{s3.2}} \label{sc}

It is useful to find that
\bea
&&\phantom{=} \a_{T\mJ} \lt( d_{T\mJ}^{j_1j_2j_3} \rt)^2+\a_{\mN} \lt( d_{\mN}^{j_1j_2j_3} \rt)^2=
 \f{2c^3}{3}\lt( \lt( d_{T\mJ}^{j_1j_2j_3} \rt)^2  +\lt( d_{T\mJ}^{j_2j_3j_1} \rt)^2  +\lt( d_{T\mJ}^{j_3j_1j_2} \rt)^2 \rt) \nn\\
&&=\f{4}{(2n)^{14}c} \lt( h_{331}^{j_1j_2j_3}+2h_{322}^{j_1j_2j_3}-12h_{222}^{j_1j_2j_3} \rt)
   -\f{8(n^2-1)}{3(2n)^{14}}h_{\a}^{j_1j_2j_3} \nn\\
&& \phantom{=}
   +\f{4(n^2-1)^2c}{9(2n)^{14}} (h_{500}^{j_1j_2j_3}-h_{400}^{j_1j_2j_3}-h_{\b}^{j_1j_2j_3}).
\eea
Here we have
\bea \label{hab}
&& h_{\a}^{j_1j_2j_3}=\f{1}{(s_{j_1j_2}s_{j_2j_3}s_{j_3j_1})^3}
                      \lt( \f{c_{j_1j_2}}{s_{j_1j_2}^5}(2s_{j_1j_2}^2-s_{j_2j_3}^2-s_{j_3j_1}^2)
                           +\f{c_{j_2j_3}}{s_{j_2j_3}^5}(2s_{j_2j_3}^2-s_{j_3j_1}^2-s_{j_1j_2}^2) \rt.\nn\\
&& \phantom{h_{\a}^{j_1j_2j_3}=\f{1}{(s_{j_1j_2}s_{j_2j_3}s_{j_3j_1})^3} ~~ }\lt.
                           +\f{c_{j_3j_1}}{s_{j_3j_1}^5}(2s_{j_3j_1}^2-s_{j_1j_2}^2-s_{j_2j_3}^2)  \rt), \nn\\
&& h_{\b}^{j_1j_2j_3}= \f{ c_{j_1j_2} c_{j_2j_3} } { s_{j_1j_2}^5 s_{j_2j_3}^5 }
                      +\f{ c_{j_2j_3} c_{j_3j_1} } { s_{j_2j_3}^5 s_{j_3j_1}^5 }
                      +\f{ c_{j_3j_1} c_{j_1j_2} } { s_{j_3j_1}^5 s_{j_1j_2}^5 },
\eea
and
\be \label{hmpq}
h_{mpq}^{j_1j_2j_3}=\f{1}{s_{j_1j_2}^{2m} s_{j_2j_3}^{2p}s_{j_3j_1}^{2q}}+cyc.
\ee
with the indexes $mpq$ being totally symmetric and $cyc.$\ being some possible cyclic terms. For examples, there is
\be
h_{mmm}^{j_1j_2j_3}=\f{1}{s_{j_1j_2}^{2m} s_{j_2j_3}^{2m}s_{j_3j_1}^{2m}},
\ee
and for $m\neq p$ there is
\be
h_{mmp}^{j_1j_2j_3}=\f{1}{s_{j_1j_2}^{2m} s_{j_2j_3}^{2m}s_{j_3j_1}^{2p}}
                   +\f{1}{s_{j_1j_2}^{2m} s_{j_2j_3}^{2p}s_{j_3j_1}^{2m}}
                   +\f{1}{s_{j_1j_2}^{2p} s_{j_2j_3}^{2m}s_{j_3j_1}^{2m}},
\ee
and for $m\neq p, ~ m \neq q, ~ p\neq q$ there is
\be
h_{mpq}^{j_1j_2j_3}=\f{1}{s_{j_1j_2}^{2m} s_{j_2j_3}^{2p}s_{j_3j_1}^{2q}}
                   +\f{1}{s_{j_1j_2}^{2m} s_{j_2j_3}^{2q}s_{j_3j_1}^{2p}}
                   +\f{1}{s_{j_1j_2}^{2q} s_{j_2j_3}^{2m}s_{j_3j_1}^{2p}}
                   +\f{1}{s_{j_1j_2}^{2p} s_{j_2j_3}^{2m}s_{j_3j_1}^{2q}}
                   +\f{1}{s_{j_1j_2}^{2p} s_{j_2j_3}^{2q}s_{j_3j_1}^{2m}}
                   +\f{1}{s_{j_1j_2}^{2q} s_{j_2j_3}^{2p}s_{j_3j_1}^{2m}}.
\ee

Another useful relation is
\bea
&&\phantom{=}
\a_{T\mK} \lt( d_{T\mK}^{j_1j_2j_3} \rt)^2+\a_{\mQ} \lt( d_{\mQ}^{j_1j_2j_3} \rt)^2+\a_{\mR} \lt( d_{\mR}^{j_1j_2j_3} \rt)^2 \nn\\
&&=\f{163c^3}{455} \lt( 11 \lt( \lt( d_{T\mJ}^{j_1j_2j_3} \rt)^2
                                    +\lt( d_{T\mJ}^{j_2j_3j_1} \rt)^2
                                    +\lt( d_{T\mJ}^{j_3j_1j_2} \rt)^2 \rt) \rt. \nn\\
&& \lt. \phantom{ \lt( d_{T\mJ}^{j_1j_2j_3} \rt)^2 }
                       -4\lt( d_{T\mJ}^{j_1j_2j_3}d_{T\mJ}^{j_2j_3j_1}
                              +d_{T\mJ}^{j_2j_3j_1}d_{T\mJ}^{j_3j_1j_2}
                              +d_{T\mJ}^{j_3j_1j_2}d_{T\mJ}^{j_1j_2j_3} \rt)
                   \rt) \\
&&=\f{1}{2129920 n^{16}c}I
   -\f{n^2-1}{6389760n^{16}}II
   +\f{(n^2-1)^2c}{230031360n^{16}}III
   -\f{(n^2-1)^4 c^2}{172523520n^{14}}IV
   +\f{(n^2-1)^6 c^3}{517570560n^{12}}, \nn
\eea
where
\bea
&& I=308 h_{440}^{j_1j_2j_3}+1445 h_{422}^{j_1j_2j_3}-120h_{332}^{j_1j_2j_3}-504h_{431}^{j_1j_2j_3}
     -16(n^2+2) \lt( 14h_{331}^{j_1j_2j_3}+11h_{322}^{j_1j_2j_3}\rt) \nn\\
&& \phantom{I=}
     +192(n^2+2)^2 h_{222}^{j_1j_2j_3}  ,\nn\\
&& II=1755 h_{511}^{j_1j_2j_3}+910h_{322}^{j_1j_2j_3}-520h_{421}^{j_1j_2j_3}
      -168(n^2+13)h_{222}^{j_1j_2j_3}-188(3n^2+13)h_{411}^{j_1j_2j_3} \nn\\
&& \phantom{II=}-4(11n^2-104)h_{321}^{j_1j_2j_3}
      -56n^2 h_{420}^{j_1j_2j_3}+32(n^2+2)(3n^2+13)h_{311}^{j_1j_2j_3}, \nn\\
&& III=8775 h_{600}^{j_1j_2j_3}+3900 h_{330}^{j_1j_2j_3}
       -1560(n^2+9)h_{500}^{j_1j_2j_3}-1560(n^2+2)h_{320}^{j_1j_2j_3} \\
&& \phantom{III=}
       +48(3n^4+26n^2+117)h_{400}^{j_1j_2j_3}
       +16(25n^4+149n^2+156)h_{220}^{j_1j_2j_3} \nn\\
&& \phantom{III=}
       +88n^2(n^2-1)h_{211}^{j_1j_2j_3}
       -192n^2(n^2-1)(n^2+2)h_{111}^{j_1j_2j_3},  \nn\\
&& IV=65 h_{300}^{j_1j_2j_3}-4(3n^2+13)h_{200}^{j_1j_2j_3}.  \nn
\eea

\section{One short interval on cylinder} \label{sd}

This appendix is a byproduct of the paper.
Here we compute the R\'enyi entropy for the case of a short interval with length $\ell$ on a cylinder. We choose the spatial part of the 2D CFT is a circle of length $L$. The R\'enyi entanglement entropy of $A$ is known \cite{Calabrese:2004eu}
\be \label{e4}
S_n=\f{c}{6} \lt( 1+\f{1}{n} \rt) \log \lt( \f{L}{\pi\e}\sin\f{\pi\ell}{L} \rt).
\ee
Using the OPE of twist operators, the result was reproduced to order $O(\ell^6/L^6)$ in \cite{Chen:2013kpa}, and the result in this paper allows us to calculate to order $O(\ell^8/L^8)$.

It is shown that in \cite{Chen:2013kpa} that the R\'enyi entropy is
\be
\Tr \r^n_A=c_n \ell^{-\f{c}{6}\lt( n-\f{1}{n} \rt)} \lt( \sum_K d_K \ell^{h_K } \lag \Phi_K(0) \rag_L \rt)^2,
\ee
with $K$ being the summation over all the linear independent holomorphic quasiprimary operators constructed solely by the operators in the Virasoro Verma module.

With the transformation formulas in the previous section we could get
\bea
&& \lag T(0) \rag_L=\f{\pi^2c}{6L^2},
~~~ \lag \mc A(0) \rag_L=\f{\pi^4c(5c+22)}{180L^4},
~~~ \lag \mc B(0) \rag_L=-\frac{62 \pi ^6 c}{525 L^6},  \nn\\
&& \lag \mc D(0) \rag_L=\f{\pi^6 c(2c-1)(5c+22)(7c+68)}{216(70c+29)L^6},
~~~ \lag \mc E(0) \rag_L=\frac{23452 \pi ^8 c}{59535 L^8},  \nn\\
&& \lag \mc H(0) \rag_L=-\frac{13 \pi ^8 c (5 c+22) (465 c-127)}{10125 (105 c+11) L^8},  \\
&& \lag \mc I(0) \rag_L=\frac{\pi ^8 c (2 c-1) (3 c+46) (5 c+3) (5 c+22) (7 c+68)}{1296 (5 c (210 c+661)-251) L^8}. \nn
\eea
Then we could find the R\'enyi entanglement entropy
\bea
&& S_n=-\f{1}{n-1}\log \Tr \r_A^n  \nn\\
&&\phantom{S_n}
=\f{c}{6} \lt( 1+\f{1}{n} \rt)
\lt( \log\f{\ell}{\e}-\frac{\pi ^2 \ell ^2}{6 L^2}-\frac{\pi ^4 \ell ^4}{180 L^4}-\frac{\pi ^6 \ell ^6}{2835 L^6}
                     -\frac{\pi ^8 \ell ^8}{37800 L^8}+{O}\left(\f{\ell ^{10}}{L^{10}}\right) \rt),
\eea
which matches (\ref{e4}) to the order of $O(\ell^8/L^8)$.

The finite temperature effect is the same with the  finite length case if we substitute $L \to i\b$ with $\b$ being the inverse temperature.

\section{Some useful summation formulas} \label{se}

In this appendix we summarize some formulas that are needed in our calculation. We define
\be \label{fm}
f_m=\sum_{j=1}^{n-1}\f{1}{ \lt( \sin\f{\pi j}{n} \rt)^{2m}}.
\ee
We have also defined $h_\a^{j_1j_2j_3}$, $h_\b^{j_1j_2j_3}$ in (\ref{hab}), $h_{mpq}^{j_1j_2j_3}$ in (\ref{hmpq}), and $g_\a^{j_1j_2j_3j_4}$, $g_\b^{j_1j_2j_3j_4}$, $g_\g^{j_1j_2j_3j_4}$, $g_\d^{j_1j_2j_3j_4}$ in (\ref{gabgd}).

Explicitly we need
\bea \label{fm18}
&& f_1=\frac{n^2-1}{3}, ~~~ f_2=\frac{(n^2-1) \left(n^2+11\right)}{45} , ~~~
f_3=\frac{(n^2-1)  \left(2 n^4+23 n^2+191\right)}{945} ,  \nn\\
&& f_4=\frac{(n^2-1) \left(n^2+11\right) \left(3 n^4+10 n^2+227\right)}{14175},  \nn\\
&& f_5=\frac{(n^2-1) \left(2 n^8+35 n^6+321 n^4+2125 n^2+14797\right)}{93555}, \\
&& f_6=\frac{(n^2-1) \left(1382 n^{10}+28682 n^8+307961 n^6+2295661 n^4+13803157 n^2+92427157\right)}{638512875}. \nn\\
&& f_7=\frac{(n^2-1) \left(60 n^{12}+1442 n^{10}+17822 n^8+151241 n^6+997801 n^4+5636617 n^2+36740617\right)}{273648375}, \nn\\
&& f_8=\frac{(n^2-1)}{488462349375}\left(10851 n^{14}+296451 n^{12}+4149467 n^{10}+39686267 n^8+292184513 n^6  \rt.  \nn\\
&&  \phantom{f_8=} \lt. +1777658113 n^4+9611679169 n^2+61430943169\right). \nn
\eea
The above formulas are useful because it often appears in the calculation that
\bea
&& \sum_{0\leq j_1 <j_2 \leq n-1}\f{1}{s^{2m}_{j_1j_2}}=\f{n}{2}f_m,  \nn\\
&& \sum_{0\leq j_1 <j_2<j_3 \leq n-1} h_{m00}^{j_1j_2j_3}=\f{n(n-2)}{2}f_m. \nn
\eea
There are also several summation formulas listed below.
\bea
&& \sum_{0\leq j_1 <j_2<j_3 \leq n-1} h_{\a}^{j_1j_2j_3}= \frac{2 n (n^2-1) (n^2-4) (n^2-9) \left(319 n^6+13566 n^4+152271 n^2+892244\right)}{638512875} ,\nn\\
&& \sum_{0\leq j_1 <j_2<j_3 \leq n-1} h_{\b}^{j_1j_2j_3}=\frac{n (n^2-1) (n^2-4)\left(5 n^6+58 n^4+325 n^2+1052\right)}{467775}, \nn\\
&& \sum_{0\leq j_1 <j_2<j_3 \leq n-1} h_{111}^{j_1j_2j_3}=\frac{n (n^2-1) (n^2-4) \left(n^2+47\right)}{2835}, \nn\\
&& \sum_{0\leq j_1 <j_2<j_3 \leq n-1} h_{211}^{j_1j_2j_3}=\frac{n (n^2-1) (n^2-4)\left(n^4+40 n^2+679\right)}{14175} ,\nn\\
&& \sum_{0\leq j_1 <j_2<j_3 \leq n-1} h_{220}^{j_1j_2j_3}=\frac{2 n (n^2-1) (n^2-4) \left(n^2+11\right) \left(n^2+19\right)}{14175} ,\nn\\
&& \sum_{0\leq j_1 <j_2<j_3 \leq n-1} h_{320}^{j_1j_2j_3}=\frac{2 n (n^2-1) (n^2-4)\left(6 n^6+173 n^4+2084 n^2+12137\right)}{467775}  ,\nn\\
&& \sum_{0\leq j_1 <j_2<j_3 \leq n-1} h_{330}^{j_1j_2j_3}=\frac{n (n^2-1)(n^2-4)}{638512875} \left(739 n^8+20075 n^6+355677 n^4   \rt. \nn\\
&&\phantom{\sum_{0\leq j_1 <j_2<j_3 \leq n-1} h^{j_1j_2j_3}=} \lt.+2953625 n^2+14813884\right) ,\nn\\
&& \sum_{0\leq j_1 <j_2<j_3 \leq n-1} h_{311}^{j_1j_2j_3}=\frac{n (n^2-1) (n^2-4)\left(3 n^6+125 n^4+1757 n^2+21155\right)}{467775} ,\nn\\
&& \sum_{0\leq j_1 <j_2<j_3 \leq n-1} h_{222}^{j_1j_2j_3}=\frac{n (n^2-1) (n^2-4) \left(19 n^8+875 n^6+22317 n^4+505625 n^2+5691964\right)}{273648375}, \nn\\
&& \sum_{0\leq j_1 <j_2<j_3 \leq n-1} h_{321}^{j_1j_2j_3}= \frac{n (n^2-1) (n^2-4)}{638512875}\left(473 n^8+18745 n^6+458199 n^4 \rt. \nn\\
&&\phantom{\sum_{0\leq j_1 <j_2<j_3 \leq n-1} h^{j_1j_2j_3}=} \lt.+6674755 n^2+65423828\right),\nn\\
&& \sum_{0\leq j_1 <j_2<j_3 \leq n-1} h_{420}^{j_1j_2j_3}= \frac{n (n^2-1) (n^2-4)}{638512875}\left(1621 n^8+50875 n^6+630273 n^4\rt. \nn\\
&&\phantom{\sum_{0\leq j_1 <j_2<j_3 \leq n-1} h^{j_1j_2j_3}=} \lt.+5624825 n^2+29980406\right),\nn\\
&& \sum_{0\leq j_1 <j_2<j_3 \leq n-1} h_{411}^{j_1j_2j_3}=\frac{n (n^2-1) (n^2-4)}{638512875}\left(404 n^8+17945 n^6+276297 n^4\rt. \nn\\
&&\phantom{\sum_{0\leq j_1 <j_2<j_3 \leq n-1} h^{j_1j_2j_3}=} \lt.+2703955 n^2+27241399\right), \nn
\eea

\bea
&& \sum_{0\leq j_1 <j_2<j_3 \leq n-1} h_{421}^{j_1j_2j_3}=\frac{n (n^2-1) (n^2-4)}{1915538625}\left(138 n^{10}+5819 n^8+146479 n^6+2091357 n^4   \rt. \nn\\
&&\phantom{\sum_{0\leq j_1 <j_2<j_3 \leq n-1} h^{j_1j_2j_3}=} \lt.+22440283 n^2+193043924\right), \nn \\
&& \sum_{0\leq j_1 <j_2<j_3 \leq n-1} h_{511}^{j_1j_2j_3}=\frac{n (n^2-1) (n^2-4)}{1915538625} \left(122 n^{10}+5798 n^8+100301 n^6+1060609 n^4   \rt. \nn\\
&&\phantom{\sum_{0\leq j_1 <j_2<j_3 \leq n-1} h^{j_1j_2j_3}=} \lt.+8570077 n^2+77354293\right) ,\nn\\
&& \sum_{0\leq j_1 <j_2<j_3 \leq n-1} h_{322}^{j_1j_2j_3}=\frac{n (n^2-1) (n^2-4) }{383107725}\left(6 n^{10}+299 n^8+8023 n^6+168477 n^4   \rt. \nn\\
&&\phantom{\sum_{0\leq j_1 <j_2<j_3 \leq n-1} h^{j_1j_2j_3}=} \lt.+2635831 n^2+26217764\right), \nn\\
&& \sum_{0\leq j_1 <j_2<j_3 \leq n-1} h_{331}^{j_1j_2j_3}=\frac{2 n (n^2-1) (n^2-4)}{1915538625}\left(30 n^{10}+1229 n^8+27865 n^6+529947 n^4   \rt. \nn\\
&&\phantom{\sum_{0\leq j_1 <j_2<j_3 \leq n-1} h^{j_1j_2j_3}=} \lt.+6100405 n^2+51401324\right), \nn\\
&& \sum_{0\leq j_1 <j_2<j_3 \leq n-1} h_{431}^{j_1j_2j_3}=\frac{2 n (n^2-1) (n^2-4) }{488462349375}\left(1477 n^{12}+63995 n^{10}+1496434 n^8+27120610 n^6   \rt. \nn\\
&&\phantom{\sum_{0\leq j_1 <j_2<j_3 \leq n-1} h^{j_1j_2j_3}=} \lt.+367070101 n^4+3500775395 n^2+26585391988\right), \nn\\
&& \sum_{0\leq j_1 <j_2<j_3 \leq n-1} h_{332}^{j_1j_2j_3}=\frac{4n (n^2-1) (n^2-4) }{488462349375}\left(128 n^{12}+7015 n^{10}+203101 n^8+4225095 n^6   \rt. \nn\\
&&\phantom{\sum_{0\leq j_1 <j_2<j_3 \leq n-1} h^{j_1j_2j_3}=} \lt.+78482039 n^4+1054880390 n^2+9748602232\right), \nn\\
&& \sum_{0\leq j_1 <j_2<j_3 \leq n-1} h_{422}^{j_1j_2j_3}=\frac{n (n^2-1) (n^2-4)}{488462349375}\left(709 n^{12}+37205 n^{10}+1040278 n^8+22228690 n^6   \rt. \nn\\
&&\phantom{\sum_{0\leq j_1 <j_2<j_3 \leq n-1} h^{j_1j_2j_3}=} \lt.+325794217 n^4+3892862105 n^2+34949076796\right), \nn
\eea

\bea
&& \sum_{0\leq j_1 <j_2<j_3 \leq n-1} h_{440}^{j_1j_2j_3}=\frac{2 n (n^2-1) (n^2-4) }{488462349375}\left(2757 n^{12}+88245 n^{10}+1511414 n^8+20490610 n^6   \rt. \nn\\
&&\phantom{\sum_{0\leq j_1 <j_2<j_3 \leq n-1} h^{j_1j_2j_3}=} \lt.+181728201 n^4+1105797145 n^2+4786765628\right), \nn
\eea

There are also the summations of four indexes.
\bea
&& \sum_{0\leq j_1 <j_2<j_3<j_4 \leq n-1} g_{\a}^{j_1j_2j_3j_4}=\frac{n (n^2-1) (n^2-4) (n^2-9) \left(n^2+119\right)}{28350}, \nn\\
&& \sum_{0\leq j_1 <j_2<j_3<j_4 \leq n-1} g_{\b}^{j_1j_2j_3j_4}=\frac{n (n-2) (n-3) (n^2-1) \left(n^2+11\right) \left(7 n^3+13 n^2+93 n+127\right)}{113400}, \nn
\eea

\bea
&& \sum_{0\leq j_1 <j_2<j_3<j_4 \leq n-1} g_{\g}^{j_1j_2j_3j_4}=\frac{n (n-3) (n^2-1) (n^2-4) \left(n^2+47\right)}{2835}, \nn\\
&& \sum_{0\leq j_1 <j_2<j_3<j_4 \leq n-1} g_{\d}^{j_1j_2j_3j_4}=\frac{n (n-2) (n-3) (n^2-1) \left(n^2+11\right)}{180}, \nn\\
&& \sum_{0\leq j_1 <j_2<j_3<j_4 \leq n-1} \lt( g_{\a}^{j_1j_2j_3j_4} \rt)^2=\frac{n (n^2-1) (n^2-4) (n^2-9)}{54273594375}\left(21 n^{10}+1994 n^8+105648 n^6   \rt. \nn\\
&&\phantom{\sum_{0\leq j_1 <j_2<j_3<j_4 \leq n-1} \lt( g_{\a}^{j_1j_2j_3j_4} \rt)^2=} \lt.+4785522 n^4+141534331 n^2+2127620484\right) ,\nn\\
&& \sum_{0\leq j_1 <j_2<j_3<j_4 \leq n-1} \lt( g_{\b}^{j_1j_2j_3j_4} \rt)^2=\frac{n (n-2) (n-3) (n^2-1)}{3907698795000}\left(21879 n^{13}+45093 n^{12}+699510 n^{11}   \rt. \nn\\
&&\phantom{\sum_{0\leq j_1 <j_2<j_3<j_4 \leq n-1} \lt( g_{\a}^{j_1j_2j_3j_4} \rt)^2=} \lt.+1522530 n^{10}+12198793 n^9+30819611 n^8+178371380 n^7   \rt. \nn\\
&&\phantom{\sum_{0\leq j_1 <j_2<j_3<j_4 \leq n-1} \lt( g_{\a}^{j_1j_2j_3j_4} \rt)^2=} \lt.+647286940 n^6+2857453977 n^5+14207989899 n^4   \rt. \nn\\
&&\phantom{\sum_{0\leq j_1 <j_2<j_3<j_4 \leq n-1} \lt( g_{\a}^{j_1j_2j_3j_4} \rt)^2=} \lt.+57188421110 n^3+224193314530 n^2+788902033351 n\rt. \nn\\
&&\phantom{\sum_{0\leq j_1 <j_2<j_3<j_4 \leq n-1} \lt( g_{\a}^{j_1j_2j_3j_4} \rt)^2=} \lt.+949713901397\right) ,\nn\\
&& \sum_{0\leq j_1 <j_2<j_3<j_4 \leq n-1} \lt( g_{\g}^{j_1j_2j_3j_4} \rt)^2=\frac{n (n-3) (n^2-1) (n^2-4)}{1915538625}\left(133 n^8+519 n^7+7682 n^6+45486 n^5   \rt. \nn\\
&&\phantom{\sum_{0\leq j_1 <j_2<j_3<j_4 \leq n-1} \lt( g_{\a}^{j_1j_2j_3j_4} \rt)^2=} \lt.+292677 n^4+1470231 n^3+7950068 n^2+35678964 n+146880640\right) ,\nn\\
&& \sum_{0\leq j_1 <j_2<j_3<j_4 \leq n-1} \lt( g_{\d}^{j_1j_2j_3j_4} \rt)^2=\frac{n (n-2) (n-3) (n^2-1) \left(n^2+11\right)}{56700}\left(3 n^4+23 n^3\rt. \nn\\
&&\phantom{\sum_{0\leq j_1 <j_2<j_3<j_4 \leq n-1} \lt( g_{\a}^{j_1j_2j_3j_4} \rt)^2=} \lt.+55 n^2+397 n+962\right), \nn\\
&& \sum_{0\leq j_1 <j_2<j_3<j_4 \leq n-1} g_{\a}^{j_1j_2j_3j_4} g_{\b}^{j_1j_2j_3j_4}=\frac{n (n^2-1) (n^2-4) (n^2-9)}{57466158750}\left(148 n^{10}+12742 n^8+399559 n^6   \rt. \nn\\
&&\phantom{\sum_{0\leq j_1 <j_2<j_3<j_4 \leq n-1} g_{\a}^{j_1j_2j_3j_4} g_{\b}^{j_1j_2j_3j_4}=} \lt.+9468311 n^4+149848193 n^2+1582095047\right) ,\nn
\eea

\bea
&& \sum_{0\leq j_1 <j_2<j_3<j_4 \leq n-1} g_{\a}^{j_1j_2j_3j_4} g_{\g}^{j_1j_2j_3j_4}=\frac{2 n (n^2-1) (n^2-4) (n^2-9)}{1915538625}\left(13 n^8+1193 n^6+49371 n^4\rt. \nn\\
&&\phantom{\sum_{0\leq j_1 <j_2<j_3<j_4 \leq n-1} g_{\a}^{j_1j_2j_3j_4} g_{\b}^{j_1j_2j_3j_4}=} \lt.+1707707 n^2+27272116\right) ,\nn\\
&& \sum_{0\leq j_1 <j_2<j_3<j_4 \leq n-1} g_{\a}^{j_1j_2j_3j_4} g_{\d}^{j_1j_2j_3j_4}=\frac{n (n^2-1) (n^2-4) (n^2-9)}{638512875}\left(587 n^6+57813 n^4+1175013 n^2\rt. \nn\\
&&\phantom{\sum_{0\leq j_1 <j_2<j_3<j_4 \leq n-1} g_{\a}^{j_1j_2j_3j_4} g_{\b}^{j_1j_2j_3j_4}=} \lt.+18724987\right), \nn
\eea

\bea
&& \sum_{0\leq j_1 <j_2<j_3<j_4 \leq n-1} g_{\b}^{j_1j_2j_3j_4} g_{\g}^{j_1j_2j_3j_4}=\frac{n (n^2-1) (n^2-4) (n^2-9) \left(n^2+39\right)}{212837625} \left(17 n^6+511 n^4\rt. \nn\\
&&\phantom{\sum_{0\leq j_1 <j_2<j_3<j_4 \leq n-1} g_{\a}^{j_1j_2j_3j_4} g_{\b}^{j_1j_2j_3j_4}=} \lt.+8323 n^2+71789\right) ,\nn\\
&& \sum_{0\leq j_1 <j_2<j_3<j_4 \leq n-1} g_{\b}^{j_1j_2j_3j_4} g_{\d}^{j_1j_2j_3j_4}=\frac{n (n-2) (n-3) (n^2-1)}{2554051500}\left(3003 n^9+10001 n^8+105060 n^7\rt. \nn\\
&&\phantom{\sum_{0\leq j_1 <j_2<j_3<j_4 \leq n-1} g_{\a}^{j_1j_2j_3j_4} g_{\b}^{j_1j_2j_3j_4}=} \lt.+407500 n^6+2141874 n^5+7267638 n^4+28886340 n^3\rt. \nn\\
&&\phantom{\sum_{0\leq j_1 <j_2<j_3<j_4 \leq n-1} g_{\a}^{j_1j_2j_3j_4} g_{\b}^{j_1j_2j_3j_4}=} \lt.+64096300 n^2+168447723 n+207636161\right) ,\nn\\
&& \sum_{0\leq j_1 <j_2<j_3<j_4 \leq n-1} g_{\g}^{j_1j_2j_3j_4} g_{\d}^{j_1j_2j_3j_4}=\frac{n (n-3) (n^2-1) (n^2-4) \left(n^2-3 n+26\right)}{467775} \left(3 n^4+14 n^3\rt. \nn\\
&&\phantom{\sum_{0\leq j_1 <j_2<j_3<j_4 \leq n-1} g_{\a}^{j_1j_2j_3j_4} g_{\b}^{j_1j_2j_3j_4}=} \lt.+104 n^2+370 n+1429\right). \nn
\eea

\end{appendix}

%\bibliographystyle{utcaps}     %%非常好，期刊，arXiv超链接
%\bibliographystyle{jhep}    %%好，arXiv超链接
%\bibliographystyle{kp}     %%好，arXiv超链接
%\nocite{*}

%\bibliographystyle{utphys}   %%非常好，期刊，arXiv超链接
%\bibliography{zbib}

\begin{thebibliography}{10}

\bibitem{Headrick:2010zt}
M.~Headrick, ``{Entanglement Renyi entropies in holographic theories},''
  \href{http://dx.doi.org/10.1103/PhysRevD.82.126010}{{\em Phys.Rev.}
  {\bfseries D82} (2010) 126010},
\href{http://arxiv.org/abs/1006.0047}{{\ttfamily arXiv:1006.0047 [hep-th]}}.
%%CITATION = ARXIV:1006.0047;%%.

\bibitem{Calabrese:2010he}
P.~Calabrese, J.~Cardy, and E.~Tonni, ``{Entanglement entropy of two disjoint
  intervals in conformal field theory II},''
  \href{http://dx.doi.org/10.1088/1742-5468/2011/01/P01021}{{\em J.Stat.Mech.}
  {\bfseries 1101} (2011) P01021},
\href{http://arxiv.org/abs/1011.5482}{{\ttfamily arXiv:1011.5482 [hep-th]}}.
%%CITATION = ARXIV:1011.5482;%%.

\bibitem{Chen:2013kpa}
B.~Chen and J.-J. Zhang, ``{On short interval expansion of Rényi entropy},''
  \href{http://dx.doi.org/10.1007/JHEP11(2013)164}{{\em JHEP} {\bfseries 1311}
  (2013) 164},
\href{http://arxiv.org/abs/1309.5453}{{\ttfamily arXiv:1309.5453 [hep-th]}}.
%%CITATION = ARXIV:1309.5453;%%.

\bibitem{nielsen2010quantum}
M.~A. Nielsen and I.~L. Chuang, {\em Quantum computation and quantum
  information}.
\newblock Cambridge university press, 2010.

\bibitem{petz2008quantum}
D.~Petz, {\em Quantum information theory and quantum statistics}.
\newblock Springer, 2008.

\bibitem{Callan:1994py}
J.~Callan, Curtis~G. and F.~Wilczek, ``{On geometric entropy},''
  \href{http://dx.doi.org/10.1016/0370-2693(94)91007-3}{{\em Phys.Lett.}
  {\bfseries B333} (1994) 55--61},
\href{http://arxiv.org/abs/hep-th/9401072}{{\ttfamily arXiv:hep-th/9401072
  [hep-th]}}.
%%CITATION = HEP-TH/9401072;%%.

\bibitem{Calabrese:2004eu}
P.~Calabrese and J.~L. Cardy, ``{Entanglement entropy and quantum field
  theory},'' \href{http://dx.doi.org/10.1088/1742-5468/2004/06/P06002}{{\em
  J.Stat.Mech.} {\bfseries 0406} (2004) P06002},
\href{http://arxiv.org/abs/hep-th/0405152}{{\ttfamily arXiv:hep-th/0405152
  [hep-th]}}.
%%CITATION = HEP-TH/0405152;%%.

\bibitem{Calabrese:2009ez}
P.~Calabrese, J.~Cardy, and E.~Tonni, ``{Entanglement entropy of two disjoint
  intervals in conformal field theory},''
  \href{http://dx.doi.org/10.1088/1742-5468/2009/11/P11001}{{\em J.Stat.Mech.}
  {\bfseries 0911} (2009) P11001},
\href{http://arxiv.org/abs/0905.2069}{{\ttfamily arXiv:0905.2069 [hep-th]}}.
%%CITATION = ARXIV:0905.2069;%%.

\bibitem{Calabrese:2009qy}
P.~Calabrese and J.~Cardy, ``{Entanglement entropy and conformal field
  theory},'' \href{http://dx.doi.org/10.1088/1751-8113/42/50/504005}{{\em
  J.Phys.} {\bfseries A42} (2009) 504005},
\href{http://arxiv.org/abs/0905.4013}{{\ttfamily arXiv:0905.4013
  [cond-mat.stat-mech]}}.
%%CITATION = ARXIV:0905.4013;%%.

\bibitem{Klebanov:2011uf}
I.~R. Klebanov, S.~S. Pufu, S.~Sachdev, and B.~R. Safdi, ``{Renyi Entropies for
  Free Field Theories},'' \href{http://dx.doi.org/10.1007/JHEP04(2012)074}{{\em
  JHEP} {\bfseries 1204} (2012) 074},
\href{http://arxiv.org/abs/1111.6290}{{\ttfamily arXiv:1111.6290 [hep-th]}}.
%%CITATION = ARXIV:1111.6290;%%.

\bibitem{Rajabpour:2011pt}
M.~Rajabpour and F.~Gliozzi, ``{Entanglement Entropy of Two Disjoint Intervals
  from Fusion Algebra of Twist Fields},''
  \href{http://dx.doi.org/10.1088/1742-5468/2012/02/P02016}{{\em J.Stat.Mech.}
  {\bfseries 1202} (2012) P02016},
\href{http://arxiv.org/abs/1112.1225}{{\ttfamily arXiv:1112.1225 [hep-th]}}.
%%CITATION = ARXIV:1112.1225;%%.

\bibitem{Cardy:2013nua}
J.~Cardy, ``{Some results on the mutual information of disjoint regions in
  higher dimensions},''
  \href{http://dx.doi.org/10.1088/1751-8113/46/28/285402}{{\em J.Phys.}
  {\bfseries A46} (2013) 285402},
\href{http://arxiv.org/abs/1304.7985}{{\ttfamily arXiv:1304.7985 [hep-th]}}.
%%CITATION = ARXIV:1304.7985;%%.

\bibitem{Coser:2013qda}
A.~Coser, L.~Tagliacozzo, and E.~Tonni, ``{On Rényi entropies of disjoint
  intervals in conformal field theory},''
\href{http://arxiv.org/abs/1309.2189}{{\ttfamily arXiv:1309.2189 [hep-th]}}.
%%CITATION = ARXIV:1309.2189;%%.

\bibitem{Datta:2013hba}
S.~Datta and J.~R. David, ``{Renyi entropies of free bosons on the torus and
  holography},''
\href{http://arxiv.org/abs/1311.1218}{{\ttfamily arXiv:1311.1218 [hep-th]}}.
%%CITATION = ARXIV:1311.1218;%%.

\bibitem{Maldacena:1997re}
J.~M. Maldacena, ``{The Large N limit of superconformal field theories and
  supergravity},'' {\em Adv.Theor.Math.Phys.} {\bfseries 2} (1998) 231--252,
\href{http://arxiv.org/abs/hep-th/9711200}{{\ttfamily arXiv:hep-th/9711200
  [hep-th]}}.
%%CITATION = HEP-TH/9711200;%%.

\bibitem{Gubser:1998bc}
S.~Gubser, I.~R. Klebanov, and A.~M. Polyakov, ``{Gauge theory correlators from
  noncritical string theory},''
  \href{http://dx.doi.org/10.1016/S0370-2693(98)00377-3}{{\em Phys.Lett.}
  {\bfseries B428} (1998) 105--114},
\href{http://arxiv.org/abs/hep-th/9802109}{{\ttfamily arXiv:hep-th/9802109
  [hep-th]}}.
%%CITATION = HEP-TH/9802109;%%.

\bibitem{Witten:1998qj}
E.~Witten, ``{Anti-de Sitter space and holography},'' {\em
  Adv.Theor.Math.Phys.} {\bfseries 2} (1998) 253--291,
\href{http://arxiv.org/abs/hep-th/9802150}{{\ttfamily arXiv:hep-th/9802150
  [hep-th]}}.
%%CITATION = HEP-TH/9802150;%%.

\bibitem{Ryu:2006bv}
S.~Ryu and T.~Takayanagi, ``{Holographic derivation of entanglement entropy
  from AdS/CFT},'' \href{http://dx.doi.org/10.1103/PhysRevLett.96.181602}{{\em
  Phys.Rev.Lett.} {\bfseries 96} (2006) 181602},
\href{http://arxiv.org/abs/hep-th/0603001}{{\ttfamily arXiv:hep-th/0603001
  [hep-th]}}.
%%CITATION = HEP-TH/0603001;%%.

\bibitem{Ryu:2006ef}
S.~Ryu and T.~Takayanagi, ``{Aspects of Holographic Entanglement Entropy},''
  \href{http://dx.doi.org/10.1088/1126-6708/2006/08/045}{{\em JHEP} {\bfseries
  0608} (2006) 045},
\href{http://arxiv.org/abs/hep-th/0605073}{{\ttfamily arXiv:hep-th/0605073
  [hep-th]}}.
%%CITATION = HEP-TH/0605073;%%.

\bibitem{Nishioka:2009un}
T.~Nishioka, S.~Ryu, and T.~Takayanagi, ``{Holographic Entanglement Entropy: An
  Overview},'' \href{http://dx.doi.org/10.1088/1751-8113/42/50/504008}{{\em
  J.Phys.} {\bfseries A42} (2009) 504008},
\href{http://arxiv.org/abs/0905.0932}{{\ttfamily arXiv:0905.0932 [hep-th]}}.
%%CITATION = ARXIV:0905.0932;%%.

\bibitem{Takayanagi:2012kg}
T.~Takayanagi, ``{Entanglement Entropy from a Holographic Viewpoint},''
  \href{http://dx.doi.org/10.1088/0264-9381/29/15/153001}{{\em
  Class.Quant.Grav.} {\bfseries 29} (2012) 153001},
\href{http://arxiv.org/abs/1204.2450}{{\ttfamily arXiv:1204.2450 [gr-qc]}}.
%%CITATION = ARXIV:1204.2450;%%.

\bibitem{Fursaev:2006ih}
D.~V. Fursaev, ``{Proof of the holographic formula for entanglement entropy},''
  \href{http://dx.doi.org/10.1088/1126-6708/2006/09/018}{{\em JHEP} {\bfseries
  0609} (2006) 018},
\href{http://arxiv.org/abs/hep-th/0606184}{{\ttfamily arXiv:hep-th/0606184
  [hep-th]}}.
%%CITATION = HEP-TH/0606184;%%.

\bibitem{Fursaev:2007sg}
D.~V. Fursaev, ``{Entanglement entropy in quantum gravity and the Plateau
  groblem},'' \href{http://dx.doi.org/10.1103/PhysRevD.77.124002}{{\em
  Phys.Rev.} {\bfseries D77} (2008) 124002},
\href{http://arxiv.org/abs/0711.1221}{{\ttfamily arXiv:0711.1221 [hep-th]}}.
%%CITATION = ARXIV:0711.1221;%%.

\bibitem{Casini:2011kv}
H.~Casini, M.~Huerta, and R.~C. Myers, ``{Towards a derivation of holographic
  entanglement entropy},''
  \href{http://dx.doi.org/10.1007/JHEP05(2011)036}{{\em JHEP} {\bfseries 1105}
  (2011) 036},
\href{http://arxiv.org/abs/1102.0440}{{\ttfamily arXiv:1102.0440 [hep-th]}}.
%%CITATION = ARXIV:1102.0440;%%.

\bibitem{Fursaev:2012mp}
D.~Fursaev, ``{Entanglement Renyi Entropies in Conformal Field Theories and
  Holography},'' \href{http://dx.doi.org/10.1007/JHEP05(2012)080}{{\em JHEP}
  {\bfseries 1205} (2012) 080},
\href{http://arxiv.org/abs/1201.1702}{{\ttfamily arXiv:1201.1702 [hep-th]}}.
%%CITATION = ARXIV:1201.1702;%%.

\bibitem{Lewkowycz:2013nqa}
A.~Lewkowycz and J.~Maldacena, ``{Generalized gravitational entropy},''
  \href{http://dx.doi.org/10.1007/JHEP08(2013)090}{{\em JHEP} {\bfseries 1308}
  (2013) 090},
\href{http://arxiv.org/abs/1304.4926}{{\ttfamily arXiv:1304.4926 [hep-th]}}.
%%CITATION = ARXIV:1304.4926;%%.

\bibitem{Faulkner:2013yia}
T.~Faulkner, ``{The Entanglement Renyi Entropies of Disjoint Intervals in
  AdS/CFT},''
\href{http://arxiv.org/abs/1303.7221}{{\ttfamily arXiv:1303.7221 [hep-th]}}.
%%CITATION = ARXIV:1303.7221;%%.

\bibitem{Hartman:2013mia}
T.~Hartman, ``{Entanglement Entropy at Large Central Charge},''
\href{http://arxiv.org/abs/1303.6955}{{\ttfamily arXiv:1303.6955 [hep-th]}}.
%%CITATION = ARXIV:1303.6955;%%.

\bibitem{Brown:1986nw}
J.~D. Brown and M.~Henneaux, ``{Central Charges in the Canonical Realization of
  Asymptotic Symmetries: An Example from Three-Dimensional Gravity},''
\href{http://dx.doi.org/10.1007/BF01211590}{{\em Commun.Math.Phys.} {\bfseries
  104} (1986) 207--226}.
%%CITATION = CMPHA,104,207;%%.

\bibitem{Barrella:2013wja}
T.~Barrella, X.~Dong, S.~A. Hartnoll, and V.~L. Martin, ``{Holographic
  entanglement beyond classical gravity},''
\href{http://arxiv.org/abs/1306.4682}{{\ttfamily arXiv:1306.4682 [hep-th]}}.
%%CITATION = ARXIV:1306.4682;%%.

\bibitem{Faulkner:2013ana}
T.~Faulkner, A.~Lewkowycz, and J.~Maldacena, ``{Quantum corrections to
  holographic entanglement entropy},''
\href{http://arxiv.org/abs/1307.2892}{{\ttfamily arXiv:1307.2892 [hep-th]}}.
%%CITATION = ARXIV:1307.2892;%%.

\bibitem{Maloney:2007ud}
A.~Maloney and E.~Witten, ``{Quantum Gravity Partition Functions in Three
  Dimensions},'' \href{http://dx.doi.org/10.1007/JHEP02(2010)029}{{\em JHEP}
  {\bfseries 1002} (2010) 029},
\href{http://arxiv.org/abs/0712.0155}{{\ttfamily arXiv:0712.0155 [hep-th]}}.
%%CITATION = ARXIV:0712.0155;%%.

\bibitem{Campoleoni:2010zq}
A.~Campoleoni, S.~Fredenhagen, S.~Pfenninger, and S.~Theisen, ``{Asymptotic
  symmetries of three-dimensional gravity coupled to higher-spin fields},''
  \href{http://dx.doi.org/10.1007/JHEP11(2010)007}{{\em JHEP} {\bfseries 1011}
  (2010) 007},
\href{http://arxiv.org/abs/1008.4744}{{\ttfamily arXiv:1008.4744 [hep-th]}}.
%%CITATION = ARXIV:1008.4744;%%.

\bibitem{Henneaux:2010xg}
M.~Henneaux and S.-J. Rey, ``{Nonlinear $W_{infinity}$ as Asymptotic Symmetry
  of Three-Dimensional Higher Spin Anti-de Sitter Gravity},''
  \href{http://dx.doi.org/10.1007/JHEP12(2010)007}{{\em JHEP} {\bfseries 1012}
  (2010) 007},
\href{http://arxiv.org/abs/1008.4579}{{\ttfamily arXiv:1008.4579 [hep-th]}}.
%%CITATION = ARXIV:1008.4579;%%.

\bibitem{Perlmutter:2013paa}
E.~Perlmutter, ``{Comments on Renyi entropy in AdS$_3$/CFT$_2$},''
\href{http://arxiv.org/abs/1312.5740}{{\ttfamily arXiv:1312.5740 [hep-th]}}.
%%CITATION = ARXIV:1312.5740;%%.

\bibitem{Chen:2012pc}
B.~Chen, J.~Long, and Y.-n. Wang, ``{Black holes in Truncated Higher Spin
  AdS$_3$ Gravity},'' \href{http://dx.doi.org/10.1007/JHEP12(2012)052}{{\em
  JHEP} {\bfseries 1212} (2012) 052},
\href{http://arxiv.org/abs/1209.6185}{{\ttfamily arXiv:1209.6185 [hep-th]}}.
%%CITATION = ARXIV:1209.6185;%%.

\bibitem{Yin:2007gv}
X.~Yin, ``{Partition Functions of Three-Dimensional Pure Gravity},'' {\em
  Commun.Num.Theor.Phys.} {\bfseries 2} (2008) 285--324,
\href{http://arxiv.org/abs/0710.2129}{{\ttfamily arXiv:0710.2129 [hep-th]}}.
%%CITATION = ARXIV:0710.2129;%%.

\bibitem{Giombi:2008vd}
S.~Giombi, A.~Maloney, and X.~Yin, ``{One-loop Partition Functions of 3D
  Gravity},'' \href{http://dx.doi.org/10.1088/1126-6708/2008/08/007}{{\em JHEP}
  {\bfseries 0808} (2008) 007},
\href{http://arxiv.org/abs/0804.1773}{{\ttfamily arXiv:0804.1773 [hep-th]}}.
%%CITATION = ARXIV:0804.1773;%%.

\bibitem{Gaberdiel:2011zw}
M.~R. Gaberdiel, R.~Gopakumar, T.~Hartman, and S.~Raju, ``{Partition Functions
  of Holographic Minimal Models},''
  \href{http://dx.doi.org/10.1007/JHEP08(2011)077}{{\em JHEP} {\bfseries 1108}
  (2011) 077},
\href{http://arxiv.org/abs/1106.1897}{{\ttfamily arXiv:1106.1897 [hep-th]}}.
%%CITATION = ARXIV:1106.1897;%%.

\bibitem{Gaberdiel:2010ar}
M.~R. Gaberdiel, R.~Gopakumar, and A.~Saha, ``{Quantum $W$-symmetry in
  $AdS_3$},'' \href{http://dx.doi.org/10.1007/JHEP02(2011)004}{{\em JHEP}
  {\bfseries 1102} (2011) 004},
\href{http://arxiv.org/abs/1009.6087}{{\ttfamily arXiv:1009.6087 [hep-th]}}.
%%CITATION = ARXIV:1009.6087;%%.

\bibitem{Gaberdiel:2012uj}
M.~R. Gaberdiel and R.~Gopakumar, ``{Minimal Model Holography},''
  \href{http://dx.doi.org/10.1088/1751-8113/46/21/214002}{{\em J.Phys.}
  {\bfseries A46} (2013) 214002},
\href{http://arxiv.org/abs/1207.6697}{{\ttfamily arXiv:1207.6697 [hep-th]}}.
%%CITATION = ARXIV:1207.6697;%%.

\bibitem{Chen:2014kja}
B.~Chen, F.-y. Song, and J.-j. Zhang, ``{Holographic R\'enyi entropy in
  AdS$_3$/LCFT$_2$ correspondence},''
\href{http://arxiv.org/abs/1401.0261}{{\ttfamily arXiv:1401.0261 [hep-th]}}.
%%CITATION = ARXIV:1401.0261;%%.

\bibitem{Ammon:2013hba}
M.~Ammon, A.~Castro, and N.~Iqbal, ``{Wilson Lines and Entanglement Entropy in
  Higher Spin Gravity},'' \href{http://dx.doi.org/10.1007/JHEP10(2013)110}{{\em
  JHEP} {\bfseries 1310} (2013) 110},
\href{http://arxiv.org/abs/1306.4338}{{\ttfamily arXiv:1306.4338 [hep-th]}}.
%%CITATION = ARXIV:1306.4338;%%.

\bibitem{deBoer:2013vca}
J.~de~Boer and J.~I. Jottar, ``{Entanglement Entropy and Higher Spin Holography
  in AdS$_3$},''
\href{http://arxiv.org/abs/1306.4347}{{\ttfamily arXiv:1306.4347 [hep-th]}}.
%%CITATION = ARXIV:1306.4347;%%.

\end{thebibliography}

\providecommand{\href}[2]{#2}\begingroup\raggedright\endgroup

\end{document}